\documentclass[preprint,11pt]{aastex}
\usepackage{epsfig}
\usepackage{rotating}      
\tightenlines

\def	\beq	{\begin{equation}}
\def	\Chandra {{\it Chandra}}
\def	\cm	{\,{\rm cm}}
\def    \cnt    {\,{\rm cnt}}
\def    \deg    {\,{\rm deg}}
\def	\eeq	{\end{equation}}
\def    \erg    {\,{\rm erg}}

\def	\Hz	{\,{\rm Hz}}\def	\keV	{\,{\rm keV}}
\def	\kpc	{\,{\rm kpc}}
\def	\ltsim	{\lesssim}		 %apj version
\def	\Mpc	{\,{\rm Mpc}}
\def    \Ms   {\,{\rm Ms}}
\def    \Ndot   {{\dot{N}}}

\def	\pc	{\,{\rm pc}}
\def    \photon {\,{\rm photon}}
\def	\ROSAT	{{\it ROSAT}}
\def	\s		{\,{\rm s}}

\def    \yr     {\,{\rm yr}}

\def    \Aeff   {A_{\rm eff}}
\def    \Ibg    {I_{\rm bg}}
\def    \jmax   {J}
\def    \kmax   {K}
\def    \max            {{\rm max}}
\def    \min            {{\rm min}}
\def    \Nbg   {N_{\rm bg}}
\def    \NH    {N_{\rm H}}
\def    \sca            {{\rm sca}}
\def    \tarr           {t_{\rm arr}}
\def	\tausca	        {\tau_{\rm sca}}
\def	\tauscaj	{\tau_{{\rm sca},j}}

\def	\thetahm	{\theta_{h,50}}

\def	\thetasm	{\theta_{s,50}}

\def    \Tobs           {T_{\rm obs}}

\countdef\decade=200
\advance\decade by \year
\countdef\hours=201
\advance\hours by \time
\divide\hours by 60
\countdef\mins=202
\advance\mins by \hours
\multiply\mins by 60
\multiply\hours by 100
\countdef\miltime=203
\advance\miltime by \hours
\advance\miltime by \time
\advance\miltime by -\mins
\begin{document}

\title{
%------------- enable for labelling preprint ---------------------------
        \vspace*{-3.0em}
        {\normalsize\rm submitted to {\it The Astrophysical Journal}}\\ 
        \vspace*{1.0em}
%-----------------------------------------------------------------------
	Direct Extragalactic Distance Determination Using X-Ray Scattering
%	\\
%	{\small DRAFT: \today\ -- please do not circulate}
	}

\author{B.T. Draine and Nicholas A. Bond		%enter authors here
	}
\affil{Princeton University Observatory, Peyton Hall, Princeton,
NJ 08544; \\
{\tt draine@astro.princeton.edu, nbond@astro.princeton.edu}}

\begin{abstract}
We examine the feasibility of using
dust-scattered X-rays 
for direct determination of
distances to nearby galaxies with bright background AGNs,
QSOs, or GRBs. We show how the \Chandra\ X-Ray Observatory could
be used to determine the distance to M31 to an 
unprecedented absolute accuracy of
$\sim 1\%$.
\end{abstract}


\section{Introduction
	\label{sec:intro}}

Distance measurement is one of the fundamental problems of astronomy.
The distances to extragalactic objects are generally based on a ``ladder'':
(1) Luminosities of nearby (typically low-luminosity) stars
are calibrated using trigonometric parallax.
(2) Luminosities of more luminous stars 
(e.g., Cepheids, 
or stars at the tip of the red giant branch)
are then determined by observing such stars in
apparent proximity to analogues of the trigonometrically-calibrated 
low luminosity stars.
(3) Distances to nearby galaxies are then determined by photometry of what are
believed to be 
analogues of the luminous stars.
Errors are contributed by difficulties of photometric calibration,
possible physical differences between the calibrated luminous stars
and the extragalactic targets, possible blending, and
errors in corrections for interstellar extinction.
Bonanos et al.\ (2003) conclude that
distances to galaxies in the Local Group are today known to no better
than 10--15\%.

Geometric methods do exist for direct determination of extragalactic
distances.
Detached eclipsing binaries (DEBs) can be used 
(Paczy\'nski 1997), with distances to individual DEBs in the LMC
and SMC claimed to be accurate to 4--14\% (Guinan et al.\ 1998; Fitzpatrick
et al.\ 2003; Harries et al.\ 2003).
Bonanos et al.\ describe a project to determine the distance to M31 using DEBs.
The ``light echo'' of fluorescent emission from the ring around
SN 1987A has been used to estimate the distance to the LMC
(Panagia et al.\ 1991; Gould \& Uza 1998 and references therein).
Sparks (1997) has discussed the use of polarization in using light echoes
to determine distances, and Xu et al.\ (1995)
discuss using light scattered by
dust to study the 3 dimensional structure of the ISM in front of SN 1987a.
VLBI observations of the 
orbital motions of H$_2$O masers in the galaxy NGC 4258 allow the distance
to be determined to an accuracy of 4\%
(Herrnstein et al.\ 1999) if it is assumed that the
transverse motions of the maser spots coincide with the physical motion of
the gas.

Using time-delayed X-rays scattered by dust grains
to estimate
astronomical distances was
originally proposed by Tr\"umper \& Sch\"onfelder (1973).  Here we
demonstrate the feasibility of determining extragalactic distances
by repeated
X-ray imaging of 
a time-varying AGN or QSO located behind a foreground galaxy.
The distant point source will be surrounded by a time-varying X-ray halo.
Measurement of the time-delay of the halo, as a function of halo angle,
allows the distance to the foreground galaxy to be determined.
Being based purely on geometry,
the method is essentially free of systematic uncertainties.
The question is whether the time delay can be accurately measured using
a realistic observing program -- this is the focus of the present study.

The method depends on the X-ray scattering properties of 
interstellar dust grains, summarized
in \S\ref{sec:x-ray scattering}.  The geometry of 
X-ray scattering by a foreground
galaxy is described in \S\ref{sec:scattering by dust in galaxy}.
In \S\ref{sec:observability} we consider some practical limits to
the use of X-ray scattering to determine extragalactic limits.
The method depends on variability of the X-ray source; 
in \S\ref{sec:X-Ray Variability of AGNs}
we review the X-ray variability of AGNs,
in \S\ref{sec:AGN density} their availability
on the sky..

In \S\ref{sec:Distance Determination} we present a general procedure for
determining the distance to a galaxy 
from repeated images of a background
AGN and a field around it.

In \S\ref{sec:M31 Distance}
we discuss an observational program using the \Chandra\ X-Ray
Observatory to determine the
distance to M31 using X-rays from 
the BL Lac object 5C 3.76.
We carry out simulations to demonstrate the method,
including photon-counting statistics and a realistic background.
The reader interested primarily in the efficacy of this method may
wish to go directly to Figure \ref{fig:M31_distance}
and Table \ref{tab:distance uncertainties}.
A 5 Ms observational campaign with \Chandra\ could
determine the distance to M31 with an absolute uncertainty of $\sim$4\%,
using only X-ray observations;
a 10 Ms campaign can reduce the uncertainty to $<1\%$.  
We discuss how ancillary observations
of CO and H~I can be used to further improve the accuracy of the method.

We consider using background AGNs to determine the distances to other 
galaxies in \S\ref{sec:other galaxies}, but conclude that M31 is the
most favorable opportunity.
In \S\ref{sec:GRBs} we assess the
feasibility of using a background gamma-ray burst to determine extragalactic
distances, but conclude that with present X-ray telescopes
this method will likely be limited to the LMC and SMC.
Use of X-ray telescopes other than {\it Chandra} is considered in 
\S\ref{sec:other telescopes}.
Our conclusions are summarized in \S\ref{sec:discussion and summary}.

\section{\label{sec:x-ray scattering}
         X-Ray Scattering by Dust}

Interstellar dust grains
scatter X-rays through small angles, as
was first pointed out by Overbeck (1965), Slysh (1969), and Hayakawa (1970). 
Because of this scattering, an image of an X-ray point source includes
a ``halo'' of X-rays that have been scattered
by dust grains along the line of sight.
First observed by Catura (1983) using the {\it Einstein} observatory,
scattered X-ray halos have since been studied by a number
of telescopes, including {\it Einstein} (Mauche \& Gorenstein 1986),
{\it ROSAT} (Predehl \& Schmitt 1995)
and {\it Chandra} (e.g., Smith, Edgar, \& Shafer 2002).

The scattered photons have a greater path length from source to
observer than the unscattered photons.  Therefore, if the source
is time-variable, there will be a time-varying halo which is
delayed relative to the observed variations in the point source.
Because, for a given halo angle, 
the time delay depends on the distance to the source,
Tr\"umper \& Sch\"onfelder (1973) proposed that observations of
time-variable scattered halos could be used to determine distances
to variable X-ray sources.
This effect has been used to estimate the distance to
Cyg X-3 (Predehl, Burwitz, Paerels, \& Tr\"umper 2000) and
to Nova Cygni 1992 (Draine \& Tan, in preparation).
Very recently, Vaughan et al.\ (2004) used the X-ray halo around GRB 031203
-- seen through dust with $A_V \approx 3$~mag -- 
to study the dust distribution toward $\ell=255\deg$, $b=-4.6\deg$.

There continue to be uncertainties about the composition, size, and
geometry of interstellar dust grains, but
observations of X-ray scattering by dust are in reasonable
agreement with grain models that are 
approximately consistent
with a broad range of observational constraints
(Draine 2003a).
For a model of interstellar dust consisting of a size distribution of
carbonaceous grains and silicate grains,
the differential scattering cross section per H nucleon at X-ray energies
can be approximated by (Draine 2003b) 
\beq
\label{eq:dsigmasca/dOmega}
\left(\frac{d\sigma_\sca}{d\Omega}\right)_{\theta_s}
\approx
\frac{\sigma_\sca}{2\pi\theta_s}\frac{d}{d\theta_s} 
\left[
\frac{(\theta_s/\thetasm)^2}
{1+(\theta_s/\thetasm)^2}
\right]
~~~,
\eeq
where $\sigma_\sca$ is the total scattering cross section
per H nucleon, $\theta_s$ is the scattering angle, and
\beq
\label{eq:thetasm}
\thetasm \approx 360\arcsec \left(\frac{\keV}{E}\right)
\eeq
is the median scattering angle for photons of energy $E$.
For this model the dust has a total scattering cross section given by
\beq
\label{eq:tau_s/A_V}
\frac{\tau_\sca}{A_V} \approx 0.15 \left(\frac{E}{\keV}\right)^{-1.8}  
~~~{\rm for}~ 0.8\keV \ltsim E \ltsim 10\keV
~~~.
\eeq 
where $A_V$ is the V band extinction, in magnitudes.
%The fraction of scattered photons with scattering angle $\theta_s < \theta$
%is
%\beq
%g(\theta) \equiv \frac{1}{\sigma_{\rm sca}}
%\int_0^\theta 2\pi \sin\theta_s d\theta_s \frac{d\sigma_{\rm sca}}{d\Omega}
%\approx 
%\frac{\left(\theta/\theta_{s,50}\right)^2}
%{1+\left(\theta/\theta_{s,50}\right)^2}
%~~~.
%\eeq

\section{\label{sec:scattering by dust in galaxy}
         Scattering by Dust in a Foreground Galaxy}

\begin{figure}[t]
\centerline{\epsfig{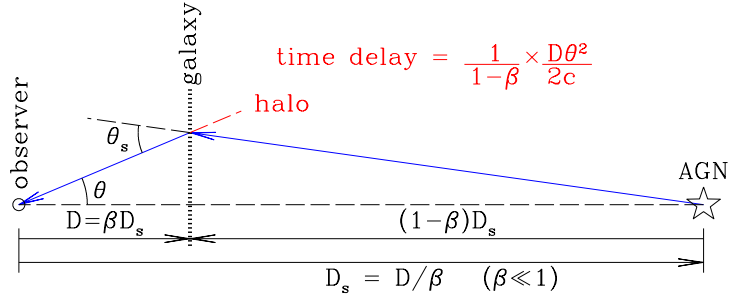}}
\caption{\label{fig:geometry}\footnotesize
       Geometry for single scattering of X-rays by dust in an
       intervening galaxy.
       }
\end{figure}

Suppose that we observe a point source with flux $F(t)$ located
behind an intervening galaxy at distance $D$ (see Fig.\ \ref{fig:geometry}).
Let the distance to the point source $D_s \gg D$.
Then X-ray photons observed at
``halo angle'' $\theta$ have
been scattered through an angle $\theta_s \approx \theta/(1-\beta)$, 
where $\beta\equiv D/D_s\ll 1$.

Consider a region in the galaxy subtending solid angle $d\Omega$,
at angular separation $\theta$ from the point source,
with H column density $\NH$.
The differential flux contributed by this region is
\beq
dH(t) = 
\left[ \frac{F(t-\delta)}{(1-\beta)^2}
\NH \left(\frac{d\sigma_\sca}{d\Omega_s}\right)_{\theta_s}
+ \Ibg\right]
d\Omega
~~~,
\eeq
where 
$\delta$ is the increased light travel time for the scattered photons,
and $\Ibg$ is background emission from unresolved sources and
diffuse emission.
For Euclidean space and $\theta\ll 1$, 
the time delay for the scattered photons is
\beq
\label{eq:tdelay}
\delta = 
\frac{1}{1-\beta}
\frac{D\theta^2}{2c}
= \frac{1.21\times10^7\s}{1-\beta}
\left(\frac{D}{\Mpc}\right)
\left(\frac{\theta}{100\arcsec}\right)^2
~~~.
\eeq

Now since $\theta_s=\theta/(1-\beta)$, we can write
\beq
\frac{d\sigma_\sca}{d\Omega} = 
\frac{(1-\beta)^2\sigma_\sca}{2\pi\theta}
\frac{d}{d\theta}
\left[
\frac{(\theta/\theta_{h,50})^2}{1+(\theta/\theta_{h,50})^2}
\right]
~~~~~~~~,~~~~~
\theta_{h,50}\equiv(1-\beta)\theta_{s,50}
~~~.
\eeq
Thus
\beq
dH(t) = \left[\frac{F(t-\delta) \NH \sigma_\sca}{2\pi\theta}
\frac{d}{d\theta}
\left[
\frac{(\theta/\theta_{h,50})^2}{1+(\theta/\theta_{h,50})^2}
\right]
+ \Ibg 
\right]
d\Omega
~~~.
\eeq

Consider annuli $j$, centered on the background point source,
with inner radii subtending angles $\psi_j$, $j=1,2,...,\jmax$.
The flux from annulus $j$ is
\begin{eqnarray}
H_j(t) &=& F(t-\delta_j)m_j^0
+ I_{{\rm bg},j}\Omega_j
~~~,
\end{eqnarray}
where
\beq
\label{eq:magnification}
m_j^0 \equiv
\tauscaj
\left[
\frac{(\psi_{j+1}/\theta_{h,50})^2}{1+(\psi_{j+1}/\theta_{h,50})^2}
-
\frac{(\psi_{j}/\theta_{h,50})^2}{1+(\psi_{j}/\theta_{h,50})^2}
\right]
\eeq
is the ``magnification'' of annulus $j$,
\beq
\tauscaj=N_{{\rm H},j}\sigma_\sca
\eeq
is the X-ray scattering optical depth averaged
over annulus $j$,
\beq
\theta_j \equiv \left(\frac{\psi_j^2+\psi_{j+1}^2}{2}\right)^{1/2}
\eeq
is the r.m.s. value of $\theta$ for the annulus,
\beq
\Omega_j=\pi(\psi_{j+1}^2-\psi_j^2)
\eeq
is the solid angle of the annulus,
and $I_{{\rm bg},j}$ is the background averaged over the annulus.
The average time delay for annulus $j$ is
\beq
\label{eq:delta_j}
\delta_j = \frac{1}{1-\beta} \frac{D}{2c}\theta_j^2
~~~.
\eeq
Note that portions of individual annuli can be ``masked''
--
for example, if there are bright foreground point sources present in
the image, or if a portion of the annulus falls beyond the detector boundary.
In such cases, the discussion below is unaffected provided only that
the scattering optical depth $\tau_{\sca,j}$ and the background
$I_{{\rm bg},j}$ averaged over the annulus
includes zero for the unusable ``masked'' regions of the annulus.
It is straightforward to continue to allow the background 
to vary from one annulus to another, but
henceforth we will assume the background $\Ibg$ to
be independent of $j$.

In the absence of other information, it is reasonable to assume
the dust to resemble Milky Way dust
(Draine 2003b),
with scattering optical depth
\beq
\tau_{\sca,j} \approx A_{V,j} \times \left(\frac{\tau_\sca}{A_V}\right)
~~~,
\eeq
Here $A_{V,j}$ is the visual extinction averaged over
annulus $j$, and $\tau_\sca/A_V$ is given in eq.\ (\ref{eq:tau_s/A_V}).

\section{Observability
         \label{sec:observability}}

Let $\Tobs$ be the time between the first and last observation.  The
largest useful halo angle is
\beq
\label{eq:Theta_max}
\Theta_\max=\left(\frac{2c\Tobs}{D}\right)^{1/2}=
91\arcsec
\left(\frac{\Tobs}{10^7\s}\right)^{1/2}
\left(\frac{\Mpc}{D}\right)^{1/2}
~~~;
\eeq
for halo angles $\theta > \Theta_\max$ the time delay exceeds $\Tobs$.
The radius
$R$ of the dusty region of the galaxy defines a second characteristic angle
\beq
\theta_R \equiv \frac{R}{D} = 1030\arcsec \left(\frac{R}{5\kpc}\right)
\left(\frac{\Mpc}{D}\right)
~~~.
\eeq
Let $\Delta\theta$ be the angular resolution of the X-ray imager.
Suppose that we wish to be able to determine the distance $D$
to within a fractional error $\epsilon$, given a sufficient number of
counts.
There are 3 different constraints which determine the maximum
distance $D$ for which this technique for distance determination
is feasible.

\begin{figure}[t]
\centerline{\epsfig{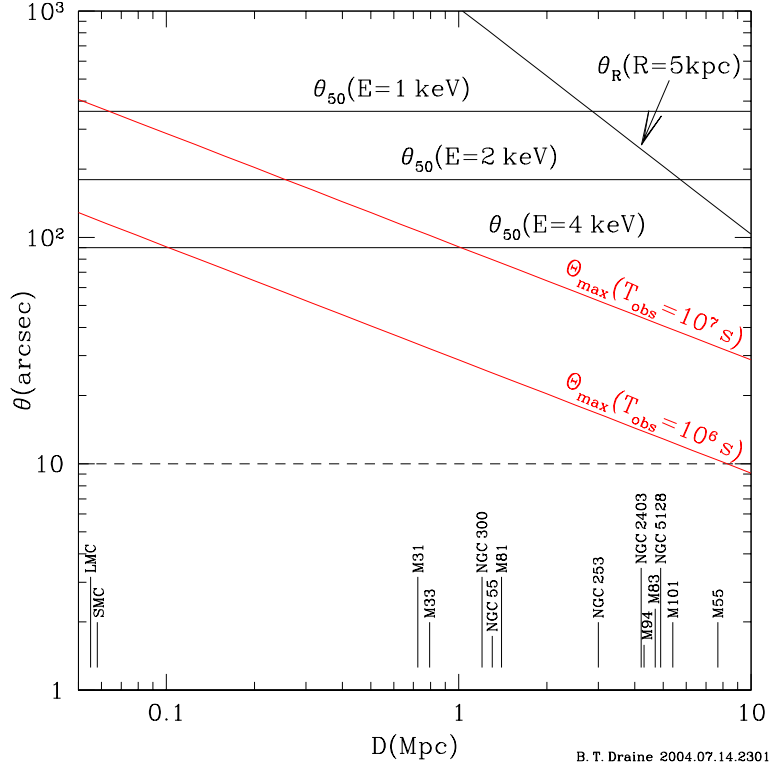}}
\caption{\label{fig:angles}\footnotesize
        Characteristic angles as a function of distance $D$:
	$\theta_R$ is the angle subtended by 5 kpc.
	$\Theta_\max(\Tobs)$ is the angle corresponding to a time
	delay $\Tobs$.
	$\theta_{50}(E)$ is the median scattering angle for photon
	energy $E$.
        }
\end{figure}

\begin{enumerate}
\item 
Even if the time delay $\delta_j$ were measured precisely for some
annulus $j$, accurate determination of the distance $D=2c\delta_j/\theta_j^2$
requires measurement of the halo angle $\theta_j$
to fractional accuracy $\epsilon/2$.
Because the distribution of dust in the galaxy will be irregular,
we cannot assume the dust to be uniformly distributed within an annulus.
Thus determination of the scattering angle to a fractional accuracy
$\epsilon/2$
requires
an
annulus width $\psi_{j+1}-\psi_j < \epsilon\theta_j$.
The annulus width is at least equal to $\Delta\theta$; thus
$\Delta\theta < \epsilon\Theta_\max$, or
\beq
D < \frac{2\epsilon^2 c\Tobs}{(\Delta\theta)^2} = 
0.827 \Mpc 
\left(\frac{\epsilon}{0.01}\right)^2
\left(\frac{\Tobs}{10^7\s}\right)
\left(\frac{1\arcsec}{\Delta\theta}\right)^2
~~~.
\eeq

\item The maximum halo angle is limited by the angular extent of the
dusty portion of the galaxy.  
Measurement of halo angles to accuracy $\epsilon$ 
would require $\Delta\theta < \epsilon\theta_R$, or
\beq
D < \frac{\epsilon R}{\Delta\theta} = 
10\Mpc 
\left(\frac{\epsilon}{0.01}\right)
\left(\frac{R}{5\kpc}\right)
\left(\frac{1\arcsec}{\Delta\theta}\right)
~~~.
\eeq

\item Suppose $\tau_\sca$ is uniform for $r < R$.
The count rate of scattered photons then varies as
\beq
H_\sca(t)=\sum_{j=1}^\jmax H_j(t) \approx
\left[
\frac{
  \left[{\rm min}(\Theta_\max,\theta_R)/\theta_{h,50}\right]^2
  }
  {
   1+\left[{\rm min}(\Theta_\max,\theta_R)/\theta_{h,50}\right]^2
  }
\right] F \tausca
~~~.
\eeq
From Figure \ref{fig:angles} we see that for 
$1\Mpc\ltsim D \ltsim 50 \Mpc$ and $\Tobs < 10^7\s$,
$\Theta_\max < \theta_R \ll \theta_{h,50}$, so that
the halo count rate $H_\sca \propto F\tausca/D$,
showing that the method is increasingly challenging as the
distance $D$ is increased.
\end{enumerate}

\section{\label{sec:X-Ray Variability of AGNs}
	X-Ray Variability of AGNs}

\begin{figure}[t]
\centerline{\epsfig{
        file=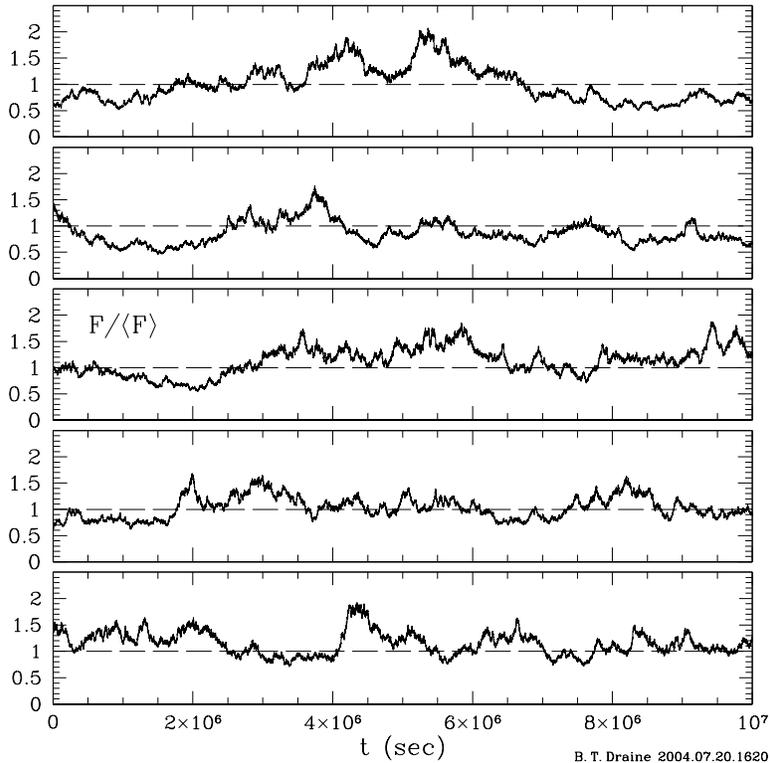,
        width=4.0in,
        angle=0}}
\caption{\label{fig:lightcurves}\footnotesize
        Flux $F$ relative to long-term average flux $\langle F\rangle$ for
        5 independent realizations of light curves with
	variability amplitude $A=0.3$ and correlation time 
	$\tau=10^6\s$, generated as described in
	Appendix \ref{app:light_curve}.
        }
\end{figure}

We simulate the X-ray light curve of an AGN with the simple 
stochastic model described
in Appendix \ref{app:light_curve}.
The model light curve is controlled by three parameters:
the average count rate $\langle F\rangle$, 
a characteristic correlation time $\tau$, and a variability amplitude
parameter $A$.
To choose realistic values of these parameters, we consider the
X-ray lightcurves obtained for AGNs using EXOSAT
(Lawrence \& Papadakis 1993) and RXTE (Markowitz et al.\ 2003)
and for QSOs using ROSAT (Ezoe et al.\ 2002).
Based on these studies, we adopt a correlation time $\tau\approx10^6\s$
as representative (see Appendix \ref{app:light_curve}).

The amplitude of variation can be characterized by
$F_{\rm var}$, the square root of the variance normalized to the
mean flux.
Six Seyfert 1 galaxies observed by Markowitz et al.\ had values of
$F_{\rm var}$
ranging from 0.22 to 0.39, corresponding to the parameter $A$
ranging from 0.22 to 0.38.
We take $A=0.3$ as typical for AGNs.
Five independent examples of synthetic light curves 
are shown in Figure \ref{fig:lightcurves}.

\section{\label{sec:AGN density}
          Density of Background AGNs}

The ROSAT North Ecliptic Pole Survey (Gioia et al.\ 2003), covering an
80.7~deg$^2$ area, detected 54 AGNs
with 0.5-2keV flux (corrected for absorption) 
greater than $2.5\times10^{-13}\erg\cm^{-2}\s^{-1}$.
The catalog is approximately consistent with
\beq
\left(\frac{dN(ES_E > S)}{d\Omega}\right)_{E=1\keV} \approx 
1.6 \left(\frac{S}{10^{-13}\erg\cm^{-2}\s^{-1}}\right)^{-3/2}~{\rm deg}^{-2}
~~~,
\eeq
where $S_E$ is the energy flux per unit photon energy $E$.
Expressed in terms of count rate $\Ndot$
for the back-illuminated CCDs of the \Chandra\ 
Advanced CCD Imaging Spectrometer
(ACIS-S BI) detector,
we estimate\footnote{%
    To estimate the \Chandra\ ACIS-S BI 
    count rates, we note that the OGLE source
    J005719.84-722533.5 with 
    $\langle ES_E\rangle_{1\keV}=1.13\times10^{-13}\erg\cm^{-2}\s^{-1}$,
    seen through $\NH=1.5\times10^{22}\cm^{-2}$ 
    has a count rate $\langle \Ndot\rangle\sim 0.041\s^{-1}$
    (Dobrzycki et al.\ 2003).
    }
\beq
\label{eq:AGN_density}
\frac{dN(\Ndot>\Ndot_{\min})}{d\Omega}
\approx 
1.6 
\left(
\frac{\Ndot_{\min}}
     {.04\cnt\s^{-1}}
\right)^{-3/2}
~{\rm deg}^{-2} ~~~.
\eeq
If we have a foreground galaxy of area $\Delta\Omega$, then we have
a 50\% probability of having at least one background source with
a \Chandra\ count rate exceeding
\beq
\label{eq:source density}
\Ndot_{\min} = 
.04 \cnt\s^{-1} \left(\frac{1.6\deg^{-2}\Delta\Omega}{\ln 2}\right)^{2/3}
= .07 \cnt\s^{-1} \left(\frac{\Delta\Omega}{\deg^2}\right)^{2/3}
~~~.
\eeq

\section{\label{sec:Distance Determination}
         Distance Determination}

Here we describe a simple method for estimating the distance from
the observed X-ray images.
Suppose that we have
$\kmax$ images; image $k$ covers ``time bin'' $k$ extending from
$t_{1,k}$ to $t_{2,k}$,
with $\Delta t_k \equiv t_{2,k} - t_{1,k}$.
The images are time-ordered, with $t_{2,k}\leq t_{1,k+1}$.
Each image is divided into circular annuli $j=1,...,J$ centered on the
background source, with the innermost annulus exterior to the p.s.f.

We now shift from discussion of fluxes to counts.
Let $N_k$ be the number of counts from the point source
in time bin $k$.
We assume that for each annulus $j$,
the photon arrival times $\tarr$ are known; if not actually known,
a random number generator is used to assign arrival times $\tarr$ for
each detected photon within each integration interval.

The time delay for annulus $j$ is $\delta_j=\alpha\theta_j^2$,
with $\alpha$ an unknown parameter to be determined.
There will be ``background'' counts due to noise in the detector,
cosmic ray events, and unresolved celestial sources.
Let $\Ibg\Aeff\Omega_j$ 
be the background count rate in the telescope in annulus $j$.
We will suppose that this background $\Ibg$ (taken here to
be uniform over the region of interest) is not known a priori, but
needs to be estimated from the observations.

For a trial value of the time delay parameter $\alpha$,
let the ``exposure fraction'' $f_{jk}(\alpha)$ 
be the fraction of the time interval 
$[t_{1,k}+\alpha\theta_j^2,t_{2,k}+\alpha\theta_j^2]$
during which the target was observed.
Obviously $f_{jk}<1$ for $t_{2,k}+\alpha\theta_j^2>t_{2,\kmax}$, 
and $f_{jk}=0$ for $t_{1,k}+\alpha\theta_j^2 \geq t_{2,\kmax}$;
$f_{jk}<1$ can also result from gaps in the observing campaign.

Let $H_{jk}(\alpha)$ be the actual number of counts in annulus $j$ 
during time interval 
$[t_{1,k}+\alpha\theta_j^2,t_{2,k}+\alpha\theta_j^2]$
(obviously, $H_{jk}=0$ when $f_{jk}=0$).
From the $H_{jk}$, we construct the total (halo + background) counts in time
bin $k$,
\beq
G_k(\alpha) = \sum_{j=1}^J H_{jk}(\alpha)
~~~.
\eeq
For a trial value of the background $\Ibg$, we use the observations
to estimate the ``magnification'' of each annulus,
\beq
\label{eq:m_j}
m_j(\alpha,\Ibg) = 
w_j(\alpha)\sum_{k=1}^K \left[ H_{jk}(\alpha) - b_{jk}(\alpha,\Ibg) \right]
~~~,
\eeq
\beq
b_{jk}(\alpha,\Ibg) \equiv \Ibg \Aeff \Omega_j f_{jk}(\alpha) \Delta t_k
~~~,
\eeq
\beq
\label{eq:w_j}
w_j(\alpha) \equiv 
\left[\sum_{k=1}^K f_{jk}(\alpha) N_{k}\right]^{-1}
~~~.
\eeq
For each trial $(\alpha,\Ibg)$ we calculate
\beq
\label{eq:chi^2}
\chi^2(\alpha,\Ibg) \equiv 
\sum_{k=1}^K
W_k(\alpha)
\frac{\left[G_k(\alpha) - M_k(\alpha,\Ibg) N_k - 
B_k(\alpha,\Ibg)\right]^2}
{\sigma_k^{2}(\alpha,\Ibg)}
~~~,
\eeq
where
\beq
B_k(\alpha,\Ibg) \equiv \sum_{j=1}^J b_{jk}
~~~,
\eeq
\beq
\label{eq:M_k}
M_k(\alpha,\Ibg) \equiv \sum_{j=1}^J m_j(\alpha,\Ibg) f_{jk}(\alpha)
~~~,
\eeq
\beq
W_k(\alpha) \equiv
\frac{G_k(\alpha)}{\sum_{\ell=1}^K G_\ell(\alpha)} 
~~~.
\eeq
The effective halo magnifications $M_k(\alpha,\Ibg)$
depend on $k$ because at early times the
outer annuli are not usable (i.e., have $f_{jk}=0$) 
since the scattered light coming from
them corresponds to the unobserved point source light curve at $t<t_{1,1}$.
The normalized weights $W_k$, with $\sum W_k = 1$, are included to
place greater weight on time bins where more counts
have been observed.

The denominator $\sigma_k^2$ in eq.\ (\ref{eq:chi^2}) is calculated
assuming photon counting statistics
(see Appendix \ref{app:statistical errors}):
\begin{eqnarray}
\sigma_k^2(\alpha,\Ibg) &=& 
  G_k(\alpha) 
+ N_k M_k^2
- 2\left(N_k^2+N_k^2 M_k + N_k M_k\right)S_k
+ \left(4N_k^2+N_k\right)S_k^2 +
\nonumber
\\
&&\left(N_k^2+N_k\right)\left(T_k+U_k\right) -2N_k V_k
\label{eq:sigma_k^2}
~~~,
\end{eqnarray}
where the estimated halo magnification $M_k(\alpha,\Ibg)$
is given by eq.\ (\ref{eq:M_k}), the weighting factors
$w_j(\alpha)$ are given by eq.\ (\ref{eq:w_j}), and
\begin{eqnarray}
S_k(\alpha,\Ibg) &\equiv& \sum_{j=1}^J m_j w_j f_{jk}^2
~~~,
\\
T_k(\alpha,\Ibg) &\equiv& \sum_{j=1}^J (w_j f_{jk})^2
\sum_{\ell=1}^K \left( m_j f_{j\ell}N_\ell  + b_{j\ell}\right)
~~~,
\\
U_k(\alpha,\Ibg) &\equiv& 
\sum_{\ell=1}^K N_\ell
\left( 
\sum_{j=1}^J m_j w_j f_{jk}f_{j\ell}
\right)^2
~~~,
\\
V_k(\alpha,\Ibg) &\equiv&
\sum_{j=1}^J w_j f_{jk} b_{jk}
~~~.
\end{eqnarray}
We require that 
$\sum_k f_{jk} N_k \gg 1$ so that the statistical analysis
of Appendix \ref{app:statistical errors} be valid.

There are $K\gg 1$ halo epochs.
We expect the function $\chi^2(\alpha,\Ibg)$ 
to have a minimum; 
let this be
at $(\alpha_\star,I_{{\rm bg}\star})$.
If the parameters $\alpha$ and $\Ibg$ are set to their true values,
then $\langle\chi^2\rangle = 1$ (see Appendix \ref{app:statistical errors}).
We adjust $\alpha$ and $\Ibg$ to minimize $\chi^2$, but have a large
number of annuli and epochs;
we expect
$\chi^2(\alpha_\star,I_{{\rm bg}\star})\approx 1$
if the only errors are due to Poisson statistics.
Our best estimate for the distance is
\beq
\label{eq:D_star}
D_\star = 2(1-\beta)c\alpha_\star
~~~.
\eeq
While we do not claim the above methodology to be the optimal distance
estimator -- there may be alternative approaches that are less sensitive
to Poisson noise -- we demonstrate below by direct simulation that the
method is capable of determining the distance for realistic data sets.

\section{\label{sec:M31 Distance}
         Determination of the Distance to M31: Simulations}

\subsection{Dusty Disk}

The dusty portion of the disk of M31 subtends a solid angle
$\sim0.8\deg^2$ (Xu \& Helou 1996; 
Schmidtobreick et al.\ 2000), corresponding to a 
%; major axis 2.25 deg = 28 kpc.
projected area $= 1.10\times10^{45}\cm^2$ for an assumed distance
$D\approx 770 \kpc$ (Freedman \& Madore 1990; van den Bergh 2000).
The total dust and gas mass have been 
estimated to be $(2.4\pm0.7)\times10^7M_\odot$ 
and
$M_{\rm H}\approx 2.5\times10^9 M_\odot$ within this disk (Xu \& Helou 1996,
corrected to $D=770\kpc$).
Thus  $N_{\rm H}=1.95\times10^{21}\cm^{-2}$.
Using $N_{\rm H}/A_V=1.87\times10^{21}\cm^{-2}$, we estimate
$A_V\approx1.0$ averaged over the $0.8\deg^2$ region.

The disk thickness of $\sim 200\pc$ is negligible compared to the distance
$D$, so we can treat the disk as a scattering sheet.
From eq.\ (\ref{eq:tdelay}) we see that a time delay of $10^7\s$ corresponds
to an angle $\theta\approx 100\arcsec$, so opposite sides of the scattering
region will differ in distance to us by 
$\Delta D \approx 2D\theta \sin i \approx 0.001 D\sin i$, where
$i$ is the inclination.  Inclination effects are therefore negligible, and
the scattering region can be idealized as a thin sheet in the plane of the
sky.

\subsection{5C 3.76}

For $\Omega=0.8\deg^2$,
eq.\ (\ref{eq:AGN_density}) would predict a 50\% probability of at least
one background source with \Chandra\ count rate $> 0.06\cnt\s^{-1}$.
Nature has provided at least 1 background AGN above this value:\footnote{
  Somewhat outside the dusty disk, the background AGN
  Mrk 957 has a ROSAT count rate of $0.076\cnt\s^{-1}$ -- about 60\%
  of 5C 3.76 -- and is known to show X-ray variability (Supper et al. 2001).
  Located just inside the D$_{25}$ countour of M31 
  ($\sim$20 kpc from the center), the dust
  extinction will be substantially
  lower than for 5C 3.76, making Mrk 957
  less attractive than 5C 3.76 as a background source for
  determination of the distance to M31.
  }
5C 3.76 [= WSTB 37W051 = RX J0040.2+4050; 
RA 00h40m13.7s, DEC 40d50m05s (J2000)] --
a compact nonthermal radio source
about 40$\arcmin$ ($\sim 9 \kpc$ projected distance)
from the center of M31.
It
was detected as an X-ray source by Einstein and \ROSAT,
and is classified as a BL Lac object (Perlman et al.\ 1996;
Laurent-Muehleisen et al.\ 1999; Nilsson et al.\ 2003), of unknown redshift.
5C 3.76 is located behind $N({\rm H~I})=1.34\times10^{21}\cm^{-2}$ of
atomic H (Dickey \& Brinks 1993).
If we assume that $\sim75\%$ of the gas is atomic, with the balance
either ionized or molecular, and assume that the dust is similar
to Milky Way dust (with $N_{\rm H}/A_V\approx 1.87\times10^{21}\cm^{-2}$), then
we estimate $A_V\approx 1.0$~mag for this region of M31.

Supper et al.\ (2001) reported a \ROSAT\ PSPC 0.5-2 keV count rate of 
$0.124\cnt\s^{-1}$, corresponding to
$ES_E(1\keV)\approx 3\times10^{-12}\erg\cm^{-2}\s^{-1}$.
5C 3.76 was observed at 16 different epochs by the \Chandra\ HRC,
with a mean count rate\footnote{
     5C 3.76 = object s1-75 
     was located 9.9 arcmin off-axis in the HRC observations.}
$0.122\cnt\s^{-1}$
(Williams et al.\ 2004).
With 5C 3.76 located behind $N({\rm HI})\approx 1.34\times10^{21}\cm^{-2}$,
the X-ray spectrum is presumably
strongly absorbed below 0.45~keV.
If the flux were in 1.5 keV photons, for which the 
ACIS-S camera has 
$\Aeff\approx600\cm^2$ (vs.\ $\sim200\cm^2$ for the HRC) the ACIS-S on-axis
count rate would be 
$\sim 0.4\cnt\s^{-1}$; we adopt this as the estimated count rate.

BL Lac-type objects are generally variable.
The X-ray flux from 5C 3.76 was in fact
observed to be variable by Williams et al.; 
the flux dropped
by a factor $\sim1.7$ between 2000-09-11 and 2000-10-12.
Given its X-ray brightness, variability, and location,
5C 3.76 is an attractive background source to use
to determine the distance to M31.
The variability spectrum of 5C 3.76 is not well-determined; we 
provisionally approximate it by our stochastic model
with $\tau=10^6\s$ and $A=0.3$ (see Figure \ref{fig:lightcurves}).

Because the redshift of 5C 3.76 is not known, we do not know the value of
$\beta\equiv D/D_s$.  However, it is probably safe to provisionally assume
a redshift $z > 0.03$, in which case
$D_s > 125\Mpc$, $\beta \ltsim 0.006$, and 
we can assume the factor 
$(1-\beta)\approx1$
in eq.\ (\ref{eq:D_star}).
Therefore even if the redshift is not known, the resulting uncertainty
in the distance determination will be less than 0.6\% provided $z > 0.03$.
A program to determine the distance to M31 using X-rays from 5C 3.76
should, however, include optical spectroscopy of
5C 3.76 to attempt to determine its redshift.

\subsection{Diffuse Background}

Diffuse emission in M31 was measured by \ROSAT\ (West et al.\ 1997).
At $\sim 40\arcmin$ from the center (along the major axis), unresolved sources
contribute a \ROSAT\ 0.5-2.0 keV count rate 
$\sim7\times10^{-8}\cnt\s^{-1}{\rm arcsec}^{-2}$.
Taking into account the greater effective area of the \Chandra\ ACIS-S 
camera (600 cm$^2$
at 1.5~keV, vs $\sim150\cm^2$ for \ROSAT\ PSPC) we estimate a \Chandra\ ACIS-S
background count rate $\sim2.8\times10^{-7}\cnt \s^{-1}{\rm ~arcsec}^{-2}$.

XMM-Newton recently 
observed diffuse X-ray emission from
a region 14$\arcmin$ -- 45$\arcmin$
NE of the nucleus of M31, along the major axis (Trudolyubov et al.\ 2004),
allowing an independent determination of the unresolved background.
At $\sim30\arcmin$ from the nucleus,
the diffuse 0.2--1.5~keV XMM count rate is 
$\sim 9.2\times10^{-7} \cnt\s^{-1} {\rm arcmin}^{-2}$ (the lowest
contour shown by Trudolyubov et al); from the radial 
variation seen by West et al.\ we estimate the count rate at 40$\arcmin$ to
be lower by a factor $\sim 0.75$.
Taking into account
the smaller effective area of Chandra\footnote{
   At 1.5 keV, Chandra ACIS-S and XMM have effective areas of $\sim600$ and 
   $\sim1300\cm^2$, respectively.}
we estimate that the diffuse emission will contribute
a \Chandra\ count rate 
$\Ibg^0 \Aeff \approx3\times10^{-7} \cnt\s^{-1}{\rm arcsec}^{-2}$,
in excellent agreement with the background estimated from the \ROSAT\
observations.

\subsection{Simulations}
\begin{figure}[t]
%\centerline{\fbox{Figure showing $m_j/m_j^0$ vs. $j$ for a simulation?}}
% f4.eps -> /u/draine/papers/xraydist/nbond/mj.ps
\centerline{
       \epsfig{
       file=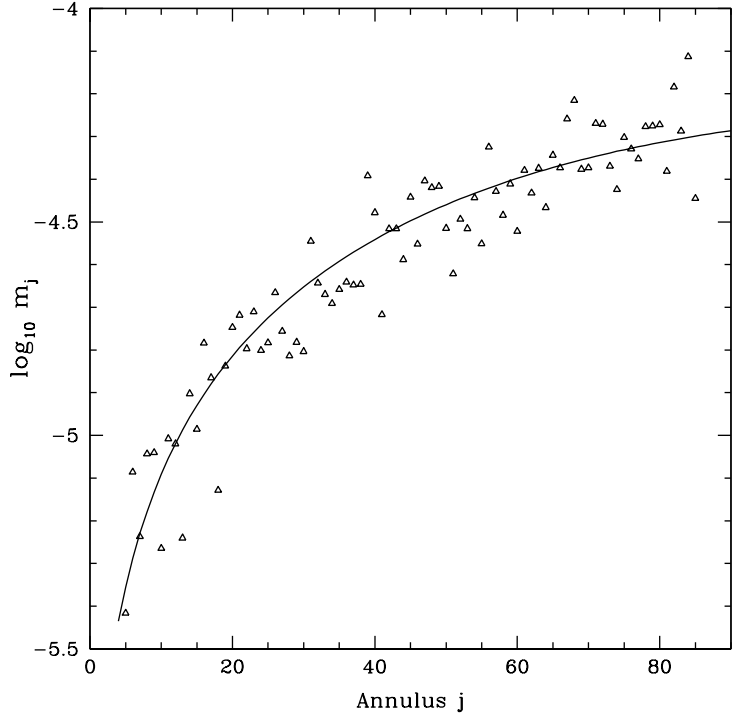,
       width=4.0in,
       angle=0}
       }
\caption{\label{fig:m_j}
       \footnotesize
       Symbols: estimated values of $m_j$ vs.\ annulus number $j$
       for a randomly-selected simulation with $\Tobs=7\times10^6\sec$,
       for the best-fit distance $D$ and background $\Ibg$.
       Solid line: $m_j^0$ used in the simulation.
       $m_j$ is largest for large $j$ because the outer annuli
       (with $\Delta\theta=1\arcsec$) have larger area.
%       Note that the fractional uncertainty in $m_j$ is largest for the
%       outermost annuli, because for the outer annuli only a small fraction
%       of the observations are usable.
       }
\end{figure}
\begin{figure}[t]
%\centerline{\fbox{Figure showing $M_k/m_k^0$ vs. $t_k$ for a simulation?}}
% f5.eps -> nbond/Mk_2.ps
\centerline{
       \epsfig{
	 file=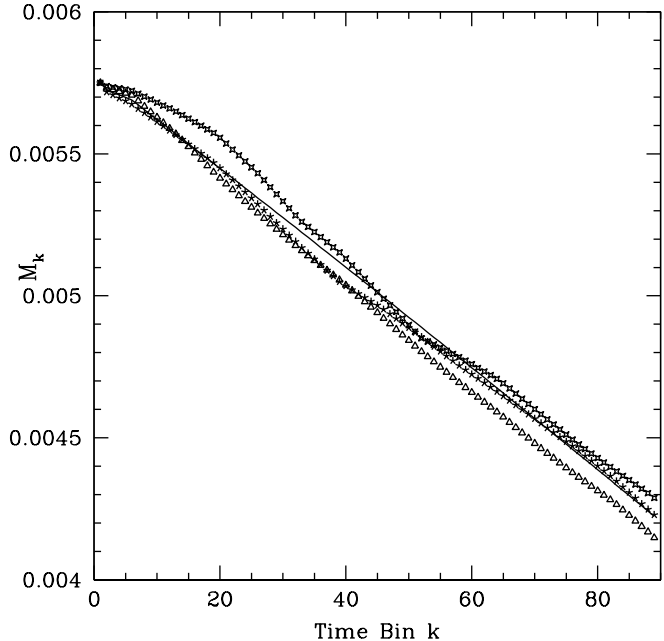,
	 width=4.0in,
	 angle=0}
       }
\caption{\label{fig:M_k}
       \footnotesize
       Effective total halo magnification $M_k$ for time bin $k$.
       Solid line: $M_k^0$ for the simulations.
       Symbols: $M_k$ for
       three randomly-selected simulations with $\Tobs=7\times10^6\s$.
       }
\end{figure}

\begin{figure}[t]
% f6.ep -> nbond/spec2.ps
\centerline{
      \epsfig{
      file=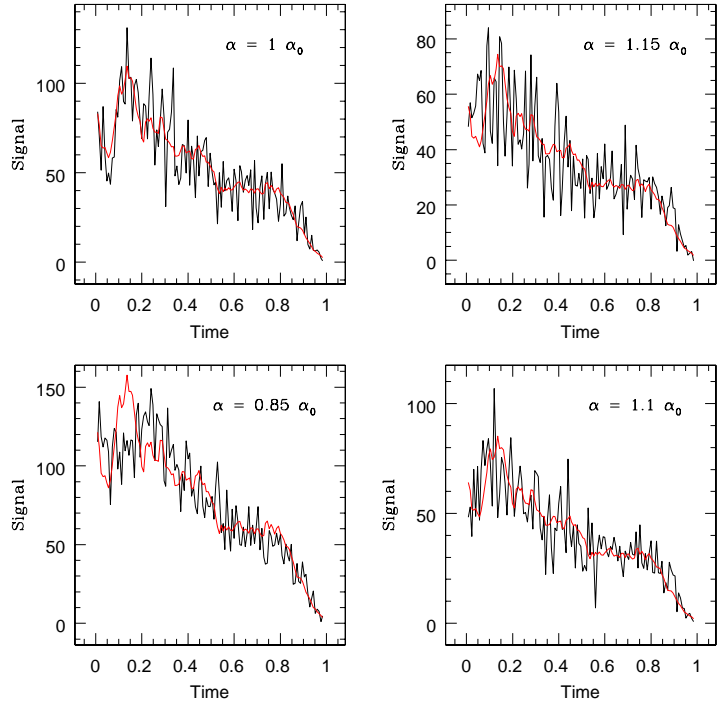,
      width=4.0in,
      angle=0}
      }
\caption{\label{fig:G_k-B_k and N_kM_k}
      \footnotesize
      $(G_k-B_k)$ (black) and $N_kM_k$ (red) for four different
      trial values of $\alpha$, for a simulation with $\Tobs=5$Ms.
      At late times $(G_k-B_k)$ and $N_kM_k$ go to zero because only
      the innermost annuli provide data.
      To the eye, $N_kM_k$ and $(G_k-B_k)$ appear to be more
      similar for the true delay $\alpha=\alpha_0$ than for
      the trial values $0.85\alpha_0$, $1.1\alpha_0$, or
      $1.15\alpha_0$, but the
      Poisson noise is large.
      }
\end{figure}

\begin{figure}[t]
% f7.eps -> nbond/bgtrials.ps
\centerline{
       \epsfig{
       file=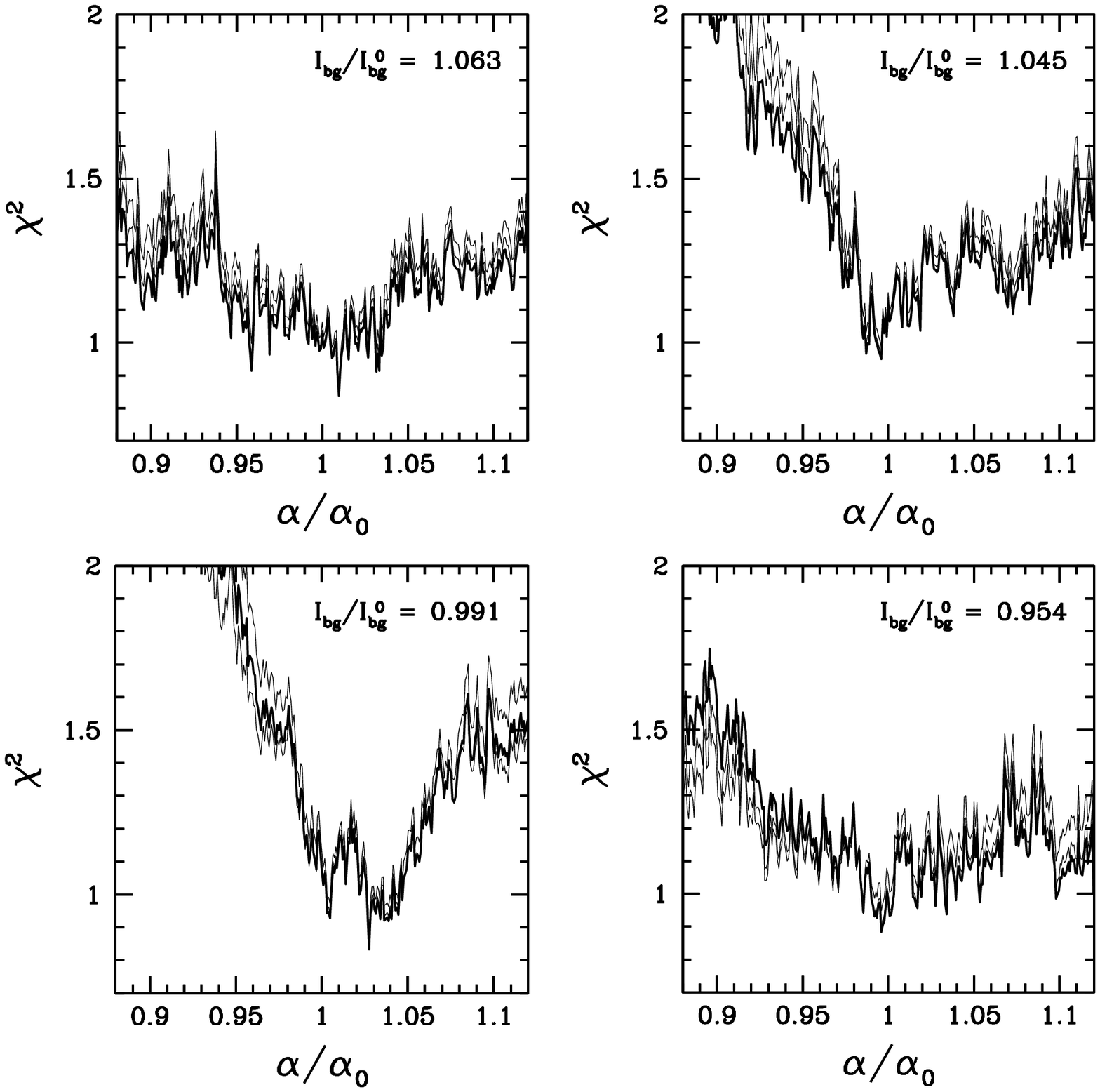,
       width=4.0in,
       angle=0}
       }
\caption{\label{fig:M31 chi^2}
       \footnotesize
       $\chi^2$ vs $D/D_0$
       for 4 randomly selected simulations with $\Tobs=7\times10^6\s$.
       In each case we show $\chi^2(\alpha)$ for
       3 trial values of $\Ibg$:
       $0.92\Ibg^0$, $1.08\Ibg^0$, and the background that
       leads to the lowest minimum of $\chi^2(\alpha)$
       For each case
       we give the estimate $I_{\rm bg,\star}$ as a fraction of $\Ibg^0$.
       }
\end{figure}

With 5C 3.76 in mind,
we simulate observations of a point source with
mean 0.5 -- 5~keV count rate $\langle \dot{N}\rangle = 0.4 \cnt\s^{-1}$,
with a stochastic light curve generated following
Appendix \ref{app:light_curve} with variability amplitude $A=0.3$ and
correlation time $\tau = 10^6\s$.
We take 
the X-rays to have a characteristic energy
$E=1.5\keV$, and approximate the disk of M31 as a uniform
scattering screen with
a scattering optical depth
$\tausca=0.072$ for the X-ray photons (corresponding to $A_V=1$~mag).
The scattering phase function for the X-rays is assumed to be
that estimated for Milky Way dust at $1.5\keV$ 
(see eq.\ \ref{eq:dsigmasca/dOmega},\ref{eq:thetasm}).\footnote{%
  While the total scattering cross section [eq.\ (\ref{eq:tau_s/A_V})]
  varies as $\sim E^{-1.8}$, the
  differential scattering cross section at 
  $\theta < \theta_{s,50}$ [see eq.\ (\ref{eq:thetasm})]
  is nearly
  independent of $E$ (see Fig.\ 8 of Draine 2003b).
  The observations considered here are at $\theta < \theta_{s,50}$,
  and thus our estimated count rates for scattered photons are insensitive
  to the assumed energy spectrum of the point source.
  }
The time delay parameter is set to $\alpha_0=D_0/(2c)$, where
$D_0=770\kpc$ [and we take $\beta=0$ in eq.\ (\ref{eq:delta_j})].

In addition to the scattered X-rays, we assume a uniform background
with a count rate $\Ibg^0 \Aeff =3\times10^{-7}\cnt\s^{-1}{\rm arcsec}^{-2}$.

For simplicity, we simulate observing campaigns without gaps,
with continuous exposures $\Tobs=$
3, 5, 7, and 10~Ms.
For each case we perform $10^3$ independent simulations.

For the 10~Ms observing campaign, the maximum useful scattering angle 
[see eq.\ (\ref{eq:Theta_max})] is
$\Theta_\max = 104\arcsec$.
For $1.5\keV$ photons, the median scattering angle is $\theta_{50}=240\arcsec$.
The mean count rate for scattered photons within $100\arcsec$ of the
point source will be 
% -------------
$\sim0.4\cdot0.072\cdot(100/240)^2/[1+(100/240)^2]=$
% -------------
$0.0043\cnt\s^{-1}$,
only $\sim$45\% of the
background count rate within this region.

We use uniform time bins of width 
$\Delta t=5\times10^4\s$, so that the point-source
counts per time bin $\langle N_k^0 \rangle \approx 2\times10^4$.  
The halo annuli are
constructed as described in Appendix \ref{app:annuli}, with minimum
annular width of $\Delta\theta=1\arcsec$, giving $J=101$ annuli 
within $104\arcsec$.

For each simulated data set, we 
construct the delay-corrected halo light curve $G_k(\alpha)$
for a trial value of the time delay parameter $\alpha$.  For different
trial values of $\Ibg$, we calculate $\chi^2(\alpha,\Ibg)$
using eq.\ (\ref{eq:chi^2}).
We repeat the procedure for different trial values of
$\alpha$, and adopt
the values $(\alpha_\star,I_{{\rm bg}\star})$ for which
$\chi^2$ is minimized.

For small values of $\Tobs$, the number of halo counts is small,
Poisson fluctuations are very substantial,
and $\alpha_\star$ may 
be far from the ``true'' value $\alpha_0$.
Because the
amount of useful data decreases as the trial value of $\alpha$ (i.e., $D$) \
is increased, Poisson noise
may result in a false minimum of $\chi^2$ for large values of $D$.
Because we have a-priori limits on the plausible range of distances
to M31, 
we search for the minimum of $\chi^2$ only over the range $[0.78D_0,1.22D_0]$.
In the event that this minimum occurs outside the range $[0.8D_0,1.2D_0]$,
we reject the distance determination altogether.
Such rejections are, however, relatively rare, especially for the longer
simulated campaigns.
In a real observing campaign, the observations
would be continued to accumulate data and eliminate such spurious
results.
For 5~Ms exposures, we had only 27 rejections in
1000 simulations; for the 10~Ms campaign, we had 0 rejections in
1000 simulations.

The method involves estimation of the annular magnifications $m_j$ from
the observations.
Figure \ref{fig:m_j} shows the $m_j$ estimated from a 7 Ms simulation.
Because of Poisson statistics, the $m_j$ are quite noisy.  However,
the $M_k$ (see eq.\ \ref{eq:M_k} and Figure \ref{fig:M_k}) are much
better behaved, as expected.

Our method seeks the time delay that minimizes the difference
between $(G_k-B_k)$ and $N_kM_k$. 
Figure \ref{fig:G_k-B_k and N_kM_k} shows
these two functions of time for one $\Tobs=5$Ms simulation,
for four trial values of $\alpha$.
To the eye, $N_kM_k$ and $(G_k-B_k)$ do appear to be more
similar for the true delay $\alpha=\alpha_0$ than for
the trial values $0.85\alpha_0$, $1.1\alpha_0$, or
$1.15\alpha_0$, but the
Poisson noise is obviously large.
Our $\chi^2$ statistic is intended to quantify the 
difference between $(G_k-B_k)$ and $N_kM_k$, where 
$\langle\chi^2\rangle = 1$ for the true time delay.

In Figure \ref{fig:M31 chi^2} we
show $\chi^2$ vs $D/D_0$ for different 
trial values of $\Ibg$, for 4 independent simulations with 
$\Tobs=7\times10^6\s$.
The value of $\Ibg$ giving the $\chi^2(\alpha)$ with the lowest minimum
ranges from $0.87\Ibg^0$ to $1.09\Ibg^0$ -- 
the background estimate is accurate to about 10\%.  
It is also apparent that the location of 
the minimum is not highly sensitive to the estimate of $\Ibg$.
Most importantly, it is seen that the minimum $\alpha_\star$ is very close
to the true value $\alpha_0$ -- within $\sim$3\% -- 
for each of these four simulations.

% f7a.eps -> nbond/3e6.ps
% f7b.eps -> nbond/5e6.ps
% f7c.eps -> nbond/7e6.ps
% f7d.eps -> nbond/1e7.ps

\begin{figure}[t]
\centerline{
       \epsfig{
       file=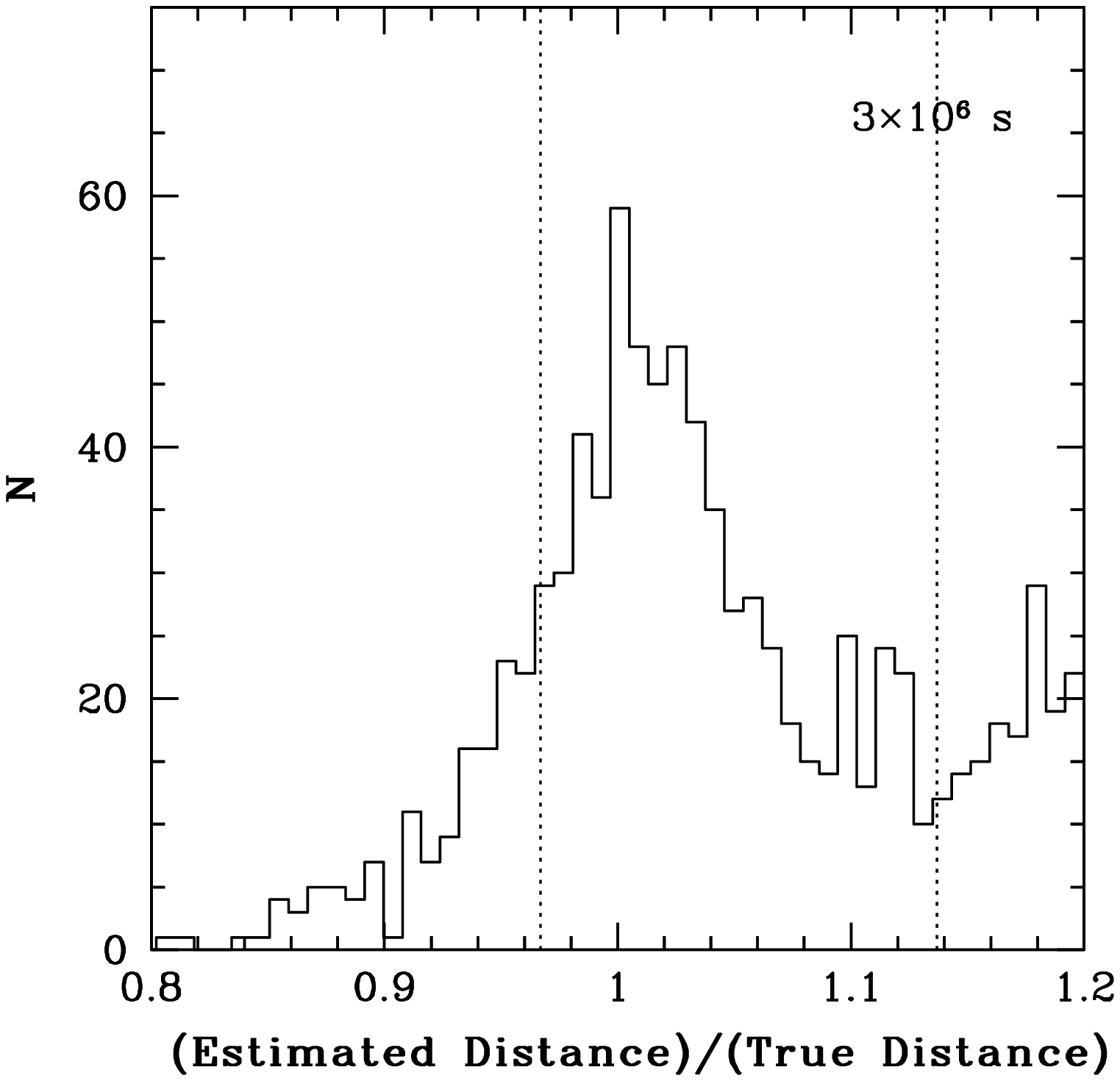,
       width=3.0in,
       angle=0}
       \epsfig{
       file=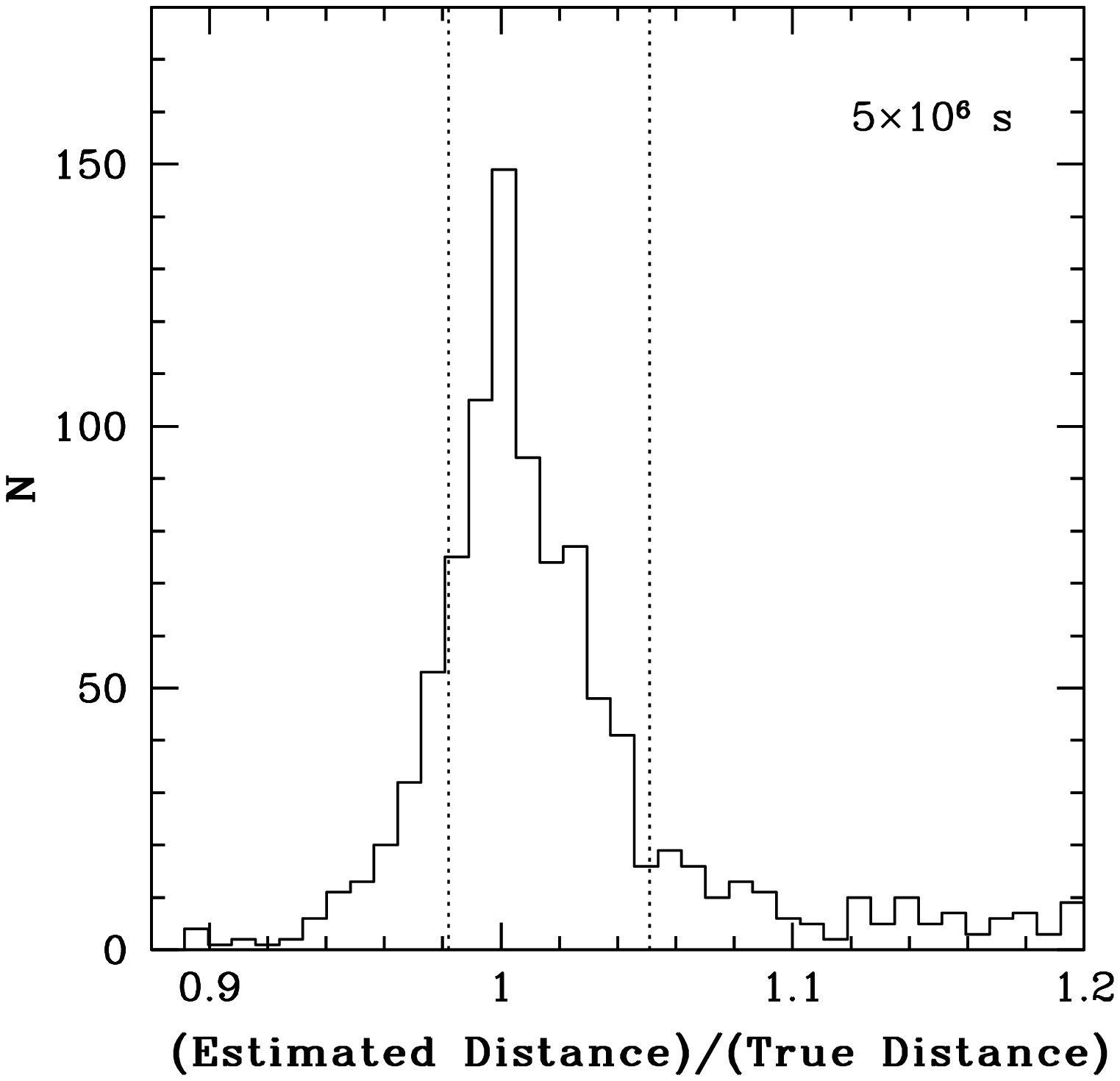,
       width=3.0in,
       angle=0}
       }

\centerline{
       \epsfig{
       file=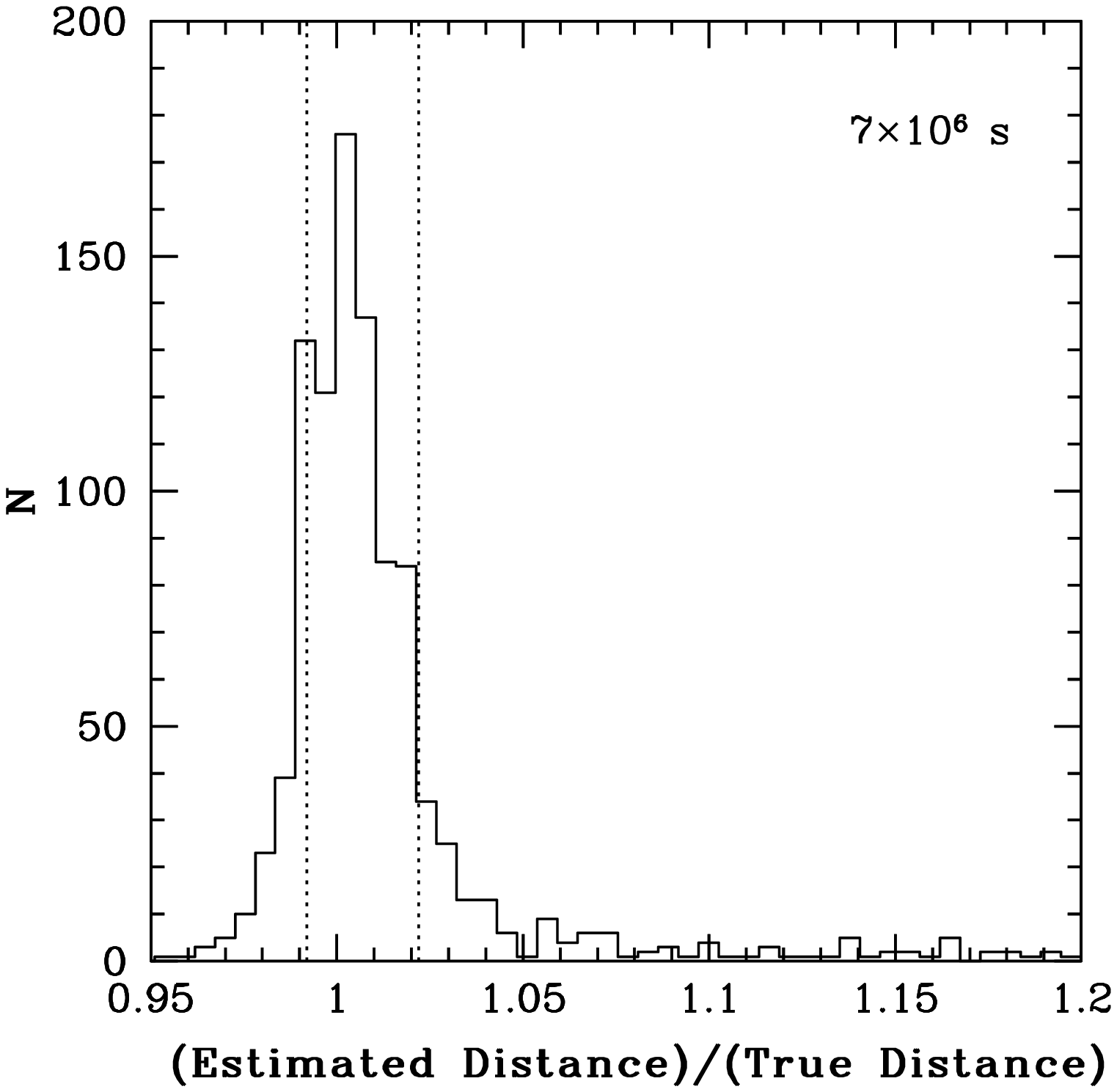,
       width=3.0in,
       angle=0}
       \epsfig{
       file=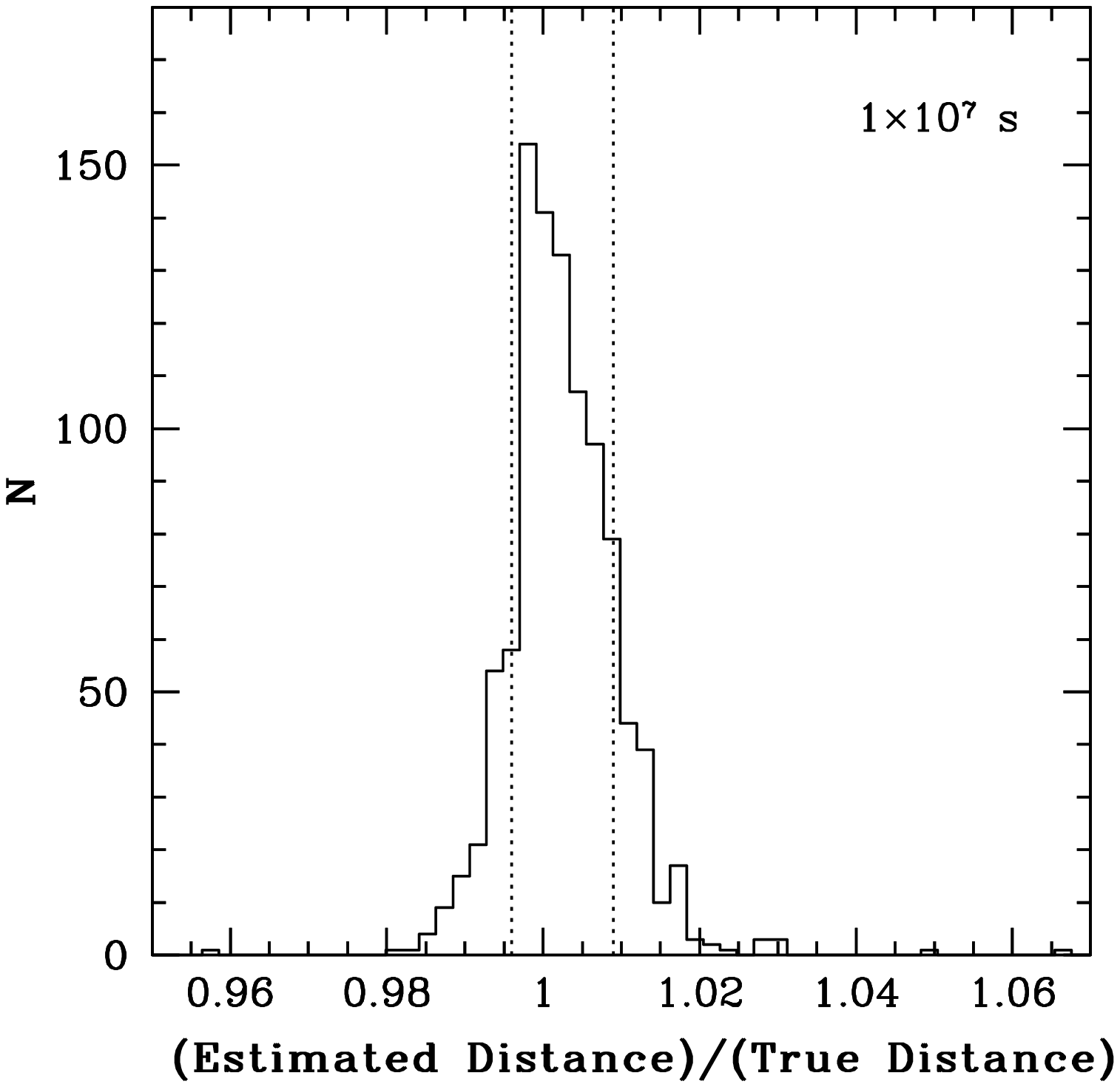,
       width=3.0in,
       angle=0}
       }

\caption{\label{fig:M31_distance}\footnotesize
       Histograms showing distribution of $D_{\rm est}/D_0$ for M31, where
       $D_{\rm est}$ is the estimated distance, for
       simulations with $T_{\rm obs}=3$, 5, 7, and 10 Ms.
       In each case, we show results for $10^3$ simulations.
       Distance estimates differing from the true distance
       by more than 20\% are rejected (see text).
       Of the remaining distance
       estimates, 68\% fall between the dotted lines, 16\% above,
       and 16\% below;
       the dotted lines can be loosely understood as $\pm1\sigma$
       uncertainties in the distance estimation.
       }
\end{figure}

Histograms of $D_\star/D_0$ are shown in Figure \ref{fig:M31_distance} for
each of the 4 values of $\Tobs$.
Table \ref{tab:distance uncertainties} gives the median, 68\% confidence
interval, and rejection fraction,
for 4 different values of $\Tobs$.
For the shortest observing campaign, $\Tobs=3\Ms$, the simulations result
in a fairly broad distribution of derived distances $D$, with about
83 of the experiments resulting in ``rejection''.
For the 917 nonrejected cases, 
the median $D/D_0 = 1.026$, and the 68\% confidence interval extends from about
$0.967 < D/D_0 < 1.137$, i.e., ``one $\sigma$'' uncertainties of 
$_{-5.9\%}^{+11.1\%}$.

As the observing time $\Tobs$ increases, the amount of useful data
increases as $\sim\Tobs^2$, because it is possible to use scattered 
halo photons at
larger
angular separations from the point source.
As a result, the accuracy of the the distance determination improves
rapidly.
As seen from Figure \ref{fig:M31_distance} 
and Table \ref{tab:distance uncertainties}, 
it appears to be possible to determine the absolute
distance to M31 to within $\pm1\%$ with a $10\Ms$ observing campaign
with \Chandra.

% 3e6:
% 
% Rejections = 83
% Average = 1.040
% Median = 1.026
% Lower Bound = 0.967
% Upper Bound = 1.137
% Width = 0.085
% 5e6:
% 
% Rejections = 27
% Average = 1.018
% Median = 1.007
% Lower Bound = 0.982
% Upper Bound = 1.051
% Width = 0.035
% 
% 7e6:
% 
% Rejections = 21
% Average = 1.012
% Median = 1.005
% Lower Bound = 0.992
% Upper Bound = 1.022
% Width = 0.015
% 
% 1e7:
% 
% Rejections = 0
% Average = 1.0027
% Median = 1.002
% Lower Bound = 0.996
% Upper Bound = 1.009
% Width = 0.0065

\begin{table}[t]
\begin{center}
\caption{\label{tab:distance uncertainties}
 Results for Simulated M31 Observing Campaigns}
\begin{tabular}{ccc}
\tableline\tableline
$\Tobs$ & $D_{\rm est}/D_0$ \tablenotemark{a}  & rejects\\
\tableline
$3\Ms$  & $1.026_{-.059}^{+.111}$     &8.3\%\\
$5\Ms$  & $1.007_{-.025}^{+.044}$ &2.7\%\\
$7\Ms$ & $1.005_{-.013}^{+.017}$  &2.1\%\\
$10\Ms$ & $1.002_{-.006}^{+.007}$  &$<0.1$\%\\
\tableline
\end{tabular}
\tablenotetext{a}{68\% confidence interval based on $10^3$ independent
simulations (see text), rejecting simulations where the distance
estimate differs from the true distance by more than 20\%.}
\end{center}
\end{table}

\subsection{\label{subsec:gaps}
            Discontinuous Observing}

Our discussion assumed, for simplicity, continuous observations of M31.
Realistic orbital and scheduling considerations would require gaps in
the observing.
The methodology presented in \S\ref{sec:Distance Determination} can be
applied to observing campaigns that include gaps.

The simplest approach would be to use only the time intervals for which
the AGN was observed, and only the counts
from annuli $j$ with arrival times $t_{\rm arr}$ such that
$t_{\rm arr}-\alpha\theta_j^2$ was a time when the AGN was observed.
This approach would, however, have the disadvantage
of discarding halo counts for which
$t_{\rm arr}-\alpha\theta_j^2$ 
falls in a gap when the AGN was not observed.

A better approach, if individual gaps are not too long, would be to 
interpolate to estimate the AGN counts $N_k$ for time intervals falling
in the gaps in the observations.
This can be accomplished using the procedure described by
Press, Rybicki, \& Hewitt (1992).
Using these interpolated counts in the same way as the actual
counts $N_k$, one could use the same procedure described above
to minimize $\chi^2(\alpha,\Ibg)$ to find the distance $D$.

Provided the interruptions in observing are sufficiently brief
($\ltsim 1$~day) that the AGN lightcurve can be reliably interpolated,
discontinuous observing should be almost as effective as
continuous observations with the same total exposure time.

\subsection{\label{sec:other wavelengths}
            Using Multiwavelength Data}

We have shown above that it is possible to determine the distance to
M31 using {\it only} X-ray imaging.  In particular, the X-ray data alone
is used to estimate the ``magnification'' $m_j$ of each
annulus, which is proportional to the amount of interstellar dust
present in the annulus.

Imaging at other wavelengths can also provide information on the
distribution of interstellar dust.  
Note that the scattering zones $j$ need not be complete annuli -- the
annuli can be subdivided in the event that the dust distribution is
nonuniform.  Indeed, if a portion of the annulus were known to be dust-free,
this should be masked off, as it would contribute 
Poisson noise from the background but no signal.
If the dust distribution over the annulus is appreciably nonuniform, it may
be advantageous to subdivide the annulus into subannuli, with low and
high dust densities.
If H~I and CO observations
are used to estimate the surface density of gas,
the magnification 
\beq \label{eq:m_j from line images}
m_j =  f(\theta) 
\left(\frac{\tausca}{\NH}\right)
\Omega_j
\left[
X_{\rm H}I_j({\rm H\,21cm}) + 2X_{\rm CO}I_j({\rm CO\,2.6mm})
\right]
\eeq
where $f(\theta)$ is the angular distribution function for X-ray
scattering\footnote{%
  With normalization $\int f(\theta)2\pi\sin\theta d\theta = (1-\beta)^{-2} 
\approx 1$.}
at halo angle $\theta$,
$\Omega_j$ is the solid angle of (sub)annulus $j$,
$I_j({\rm line})$
is the line intensity in a spectral line averaged over (sub)annulus $j$,
$X_{\rm H}$ is the conversion factor from 21cm line intensity
to H~I column density,
and
$X_{\rm CO}$ is the conversion factor
from CO line intensity to H$_2$ column density.

There are at least two different ways in which
this information could be used:
\begin{enumerate}
\item
If the angular resolution and signal/noise ratio 
of the 21cm and CO images is sufficient, and a good estimate for
$f(\theta)$ is in hand, we could use 
eq.(\ref{eq:m_j from line images})
[rather than eq.\ (\ref{eq:m_j})]
to estimate the $m_j$ for the (sub)annuli, but with the scaling factor
$\tausca/N_{\rm H}$ to be determined.
Now instead of trial values of $\alpha$ and $\Ibg$,
we would instead consider trial values of $\alpha$ and $\tausca/\NH$;
for each $(\alpha,\tausca/\NH)$ pair, 
the background $\Ibg$ is
estimated from
\beq
\Ibg(\alpha,\tausca/\NH) = 
\frac{\sum_{k=1}^K\left(G_k(\alpha) - N_k\sum_{j=1}^J m_j f_{jk}\right)}
{\Aeff \sum_{k=1}^K N_k \sum_{j=1}^J\Omega_j f_{jk}}
\eeq
Using this value of $\Ibg$, 
$\chi^2(\alpha,\tausca/\NH)$ is then calculated for the 
trial values of $\alpha$ and $\tausca/\NH$.
The best estimates of $\alpha$ and $\tausca/\NH$ are then found by
minimizing $\chi^2(\alpha,\tausca/\NH)$.

\item
Alternately, the magnifications $m_j(\alpha,\Ibg)$ 
estimated from the X-ray data
alone [eq.\ (\ref{eq:m_j})] 
could be compared to the $m_j$ estimated from 
eq.\ (\ref{eq:m_j from line images}), by computing an
error function
\begin{eqnarray}
\chi_d^2 &\equiv& \sum_j 
\frac{
\left[
m_j({\rm eq.}\ref{eq:m_j})-(\tausca/\NH)s_j)
\right]^2
}{ 
\sigma^2(m_j)
}
\\
s_j &\equiv& 
f(\theta) 
\Omega_j
\left[
X_{\rm H}I_j({\rm H\,21cm}) + 2X_{\rm CO}I_j({\rm CO\,2.6mm})
\right]
\\
\frac{\tausca}{\NH} 
&=& 
\frac{
\sum_{j=1}^J m_j(\alpha,\Ibg)s_j/\sigma^2(m_j)
}{
\sum_{j=1}^J s_j^2/\sigma_j^2(m_j)
}
~~~;
\end{eqnarray}
$\sigma^2(m_j)$ is
the variance in $m_j$ due to Poisson statistics
[see eq.\,(\ref{eq:m_j dispersion})]:
\beq
\sigma^2(m_j) \approx 
m_j \left(w_j\right)^2
\sum_k f_{jk}N_k
+
(w_j)^2 \sum_k b_{jk}
+
\left(m_jw_j\right)^2 \sum_k f_{jk}^2 N_k
~~~.
\eeq

For the wrong estimate of the distance, the estimated magnifications $m_j$
will be systematically in error.
Thus $\chi_d^2(\alpha)$ should have a minimum at the true
value of the time delay parameter $\alpha_0$
\end{enumerate}
For method (1) to be useful, it is necessary to have imaging data at
angular resolutions smaller than the larger of (a) the 
annular widths ($1\arcsec$),
or (b) the angular scale over which the dust surface density varies.
It would be useful to have 21cm and CO interferometric imaging of
this $4\arcmin\times4\arcmin$ region of M31.
In principle, the dust could be observed directly using
SCUBA on the JCMT (HPBW of
$7.5\arcsec$ at 450$\mu$m), or MIPS on the Spitzer Space Telescope 
($\sim30\arcsec$ at 160$\mu$m), but the angular resolution of these
instruments is probably insufficient to trace the dust variation from
annulus to annulus.

The best use of multiwavelength data will depend on its angular
resolution and signal/noise ratio.
We do not undertake any such simulations here.
However, it is clear that this additional information can only improve the
distance determinations.
If an observing campaign is undertaken with \Chandra, use of 
multiwavelength data should be further investigated to
make best use of the \Chandra\ observing time.

\section{\label{sec:other galaxies}
         Other Galaxies: LMC, SMC, and M81}

%\subsection{LMC and SMC}

With an H~I mass $3.1\pm0.6\times10^8M_\odot$ (Luks \& Rohlfs 1992), the
LMC has $\langle N_{\rm H}\rangle = 1\times10^{21}\cm^{-2}$ over
$\Omega=40\deg^2$.
The dust/gas ratio in the LMC is $\sim40\%$ of the Milky Way value,
hence we estimate $\langle A_V\rangle \approx 0.21$~mag, and
$\tausca \approx 0.015(E/1.5\keV)^{-1.8}$.
What is required is a bright background source.
Dobrzycki et al.\ (2003) report 5 new X-ray QSOs behind the SMC,
the brightest of which has a \Chandra\ count rate $0.041\cnt\s^{-1}$.
Over the $40\deg^2$ area of the LMC, eq.\ (\ref{eq:source density}) leads us
to expect the brightest background AGN to have
$ES_E \approx 2\times10^{-12}\erg\cm^{-2}\s^{-1}$,
with a \Chandra\ count rate $\dot{N}\approx 0.8\cnt\s^{-1}$.  This is only
twice the count rate for 5C 3.79, and $\tausca$ is likely to be a factor
of two lower than for 5C 3.79, so the actual halo brightness for an AGN
behind the LMC may not be higher than for 5C 3.79 behind M31.
The reduced distance ($D=66\kpc$ for the LMC) 
allows observation of the time-delayed
halo out to $\sim112\arcsec(\Tobs/10^6\sec)^2$, but an accurate distance
determination would still require a substantial observing campaign, and
ultimately would be limited by the uncertain three-dimensional
geometry of the LMC.
In many ways M31 is a more attractive candidate for X-ray distance
determination using AGNs.

%\subsection{M81}

As discussed in \S\ref{sec:observability}, distance determination using
X-ray halos can be applied to galaxies beyond M31, but with increasing
difficulty.  M81, at a distance $D=1.4\Mpc$, is perhaps the next most
likely candidate.  Immler \& Wang (2001) present ROSAT observations
of M81.  Aside from the nucleus of M31, 
the brightest source in the field is 
source X9, located 12.2 arcmin E of the nucleus (5 kpc projected separation).
Variable (with an amplitude exceeding a factor of 2.5;
Immler \& Wang 2001),
with a ROSAT PSPC count rate of 
$0.20\cnt\s^{-1}$,
and located in a region of M81 with 
$N({\rm H~I})\approx 3\times10^{21}\cm^{-2}$ (Immler \& Wang 2001),
X9 would be
suitable for distance determination if it were a distant background source,
but Wang (2002) argues that X9 is an intermediate-mass black hole
associated with M81.
At this time there are no known bright background sources that would
allow X-ray distance determination to M81 with existing observational
capabilities.

\section{\label{sec:GRBs}
         Using X-Rays from Gamma-Ray Bursts}

Suppose a gamma-ray burst (GRB) occurs behind a nearby galaxy
(e.g., LMC, SMC, or M31), with fluence $F_0$ in 
1--2$\keV$ X-rays.
We approximate the GRB X-rays as a single short-duration pulse.
The galaxy is approximated as a uniform thin screen
with X-ray scattering optical depth $\tausca$.  The dust is assumed
to have the scattering phase function of Milky Way dust for $1.5\keV$
X-rays [see eq.\ (\ref{eq:dsigmasca/dOmega},\ref{eq:thetasm})].

The fluence in scattered X-rays is
\begin{eqnarray}
dF_{\rm sca} &=& F_0 \tausca 
\frac{1}{\sigma_\sca}\frac{d\sigma_\sca}{d\Omega} 
2\pi \theta d\theta 
\\
&\approx& 
F_0 \tausca 
\frac{d(\theta/\thetahm)^2}{[1+(\theta/\thetahm)^2]^2}
~~~/
\end{eqnarray}
%\begin{eqnarray}
%F_{\rm sca}(<\theta) &\approx& F_0 \tausca 
%\frac{(\theta/\thetahm)^2}
%{1+(\theta/\thetahm)^2}
%\end{eqnarray}
Since the geometric time delay $t = \alpha\theta^2$, we find
\begin{equation}
dF_{\rm sca}
% &=& 
%F_0 \tausca \frac{\pi}{\sigma_\sca}
%\frac{d\sigma_\sca}{d\Omega} \frac{dt}{\alpha}
%\\
= F_0\tausca \frac{dt/t_{\rm h,50}}{[1+t/t_{\rm h,50}]^2}
~~~,
\end{equation}
where
\beq
t_{\rm h,50} = \alpha \thetahm^2 = 3.9\times10^6\s 
\left(\frac{D}{56\kpc} \right) ~~~.
\eeq
We now suppose that we image over the interval 
$(t-\Delta t/2,t+\Delta t/2)$.
The scattered photons will be observed as a ring with radius
$\theta = (t/\alpha)^{1/2}$ and width 
$(\Delta\theta)_{\rm obs} = \max[\Delta \theta,\Delta t/(2\alpha\theta)]$,
where $\Delta \theta$ is the angular resolution of the X-ray imager.
The number of scattered photons counted in the ring
will be
\begin{eqnarray}
N_r &\approx& 
F_0 \tausca \Aeff \frac{\Delta t/t_{\rm h,50}}
{(1+t/t_{\rm h,50})^2}
\\
&\approx& 23. 
\left(\frac{F_0}{10^2\cm^{-2}}\right)
\left(\frac{\tausca}{0.015}\right)
\left(\frac{56\kpc}{D}\right)
\left(\frac{240\arcsec}{\thetahm}\right)^2
\left(\frac{\Aeff}{600\cm^2}\right)
\frac{\Delta t/10^5\s}{(1+t/t_h)^2}
~~~,
\label{eq:N_r for GRB}
\end{eqnarray}
where $\Aeff$ is the effective area of the
X-ray imager.

The ring will therefore be detected with a signal-to-noise ratio
\beq
S/N \approx 
\frac{N_r}{(N_r + \Nbg)^{1/2}}
~~~,
\eeq
where the number of background counts in the ring is
\beq
\Nbg =\Ibg \Aeff 2\pi\theta(\Delta\theta)_{\rm obs}\Delta t
~~~.
\eeq

As an example, consider a GRB with a
$0.5-5\keV$ X-ray fluence $F_0 = 10^3 \photon \cm^{-2}$, 
located behind a region of
the LMC ($D=56\kpc$)
with $A_V=0.2$ ($\tausca \approx 0.015$).
A $\Delta t=2\times10^4\s$ exposure with the \Chandra\ ACIS ($A=600\cm^2$),
taken at a time $t=10^6\s$ (when the ring radius is $\theta=122\arcsec$)
will have 
$\Delta \theta = 1.22\arcsec$, and will detect $N_r=29$ counts due to
the scattered halo.
If $\Ibg\Aeff=3\times10^{-7}\cnt\s^{-1}{\rm arcsec}^{-2}$, 
there will be $\Nbg\approx 6$ background counts in the
annulus, and the ring will be detected with a signal-to-noise
ratio $S/N\approx 5$. With a width $\Delta\theta=1.22\arcsec$, the average
ring radius, and therefore the distance $D$, 
could be determined to better than 1\%.

The above example assumed a relatively high X-ray fluence
$F_0=10^3\photon\cm^{-2}$; if the fluence is lowered to 
$F_0=10^2\photon\cm^{-2}$,
the ring would have a signal-to-noise ratio of only 0.5 in a single
$\Delta t=2\times10^4\s$ exposure.  However, by taking, say, 50 such
exposures in succession and appropriately stacking them
to allow for expansion of the ring from
one exposure to the next, one could increase the S/N ratio to
$0.5\sqrt{50}=3.5$.  RXTE, with a FOV of 1130~deg$^2$, detected 8 GRBs
in the last 9 months of 1996 (Smith et al.\ 2002) corresponding to an
all-sky event rate of $\sim320\yr^{-1}$.  The detected sources had a
median 1.5-12 keV fluence of $\sim 10^2\photon\cm^{-2}$, so we estimate an
all-sky event rate of $\sim 10^2\yr^{-1}$ of GRBs with $F_0(0.5-5\keV)
> 10^2\photon\cm^{-2}$.  Since the LMC has a projected area of $40\deg^2$,
the probability per unit time of a GRB with $F_0(0.5-5\keV) >
10^2\photon\cm^{-2}$ behind the LMC is $\sim 0.1\yr^{-1}$.

While GRBs could also be used for distance determination for M31 and other
spirals, the ring counts $N_r \propto 1/D$ (see eq.\ (\ref{eq:N_r for GRB}), 
and the much smaller angular size
(0.8~deg$^2$ for M31, vs. 40~deg$^2$ for the LMC)
makes it very unlikely to have a suitably bright GRB behind M31.
Distance determination using X-ray halos around GRBs will likely be limited
to the LMC and SMC.

\section{\label{sec:other telescopes}
         Other Telescopes}

XMM-Newton offers a larger aperture than \Chandra, but its
relatively poor angular resolution ($\Delta\theta\approx6\arcsec$)
makes it unsuitable for
precision distance determination using scattered X-rays.
% $A_{\rm eff}=1304\cm^2$ @ 1.5 keV ($1800\arcsec$ FOV)
The proposed Constellation X mission
is projected to provide an order-of-magnitude
increase in collecting area ($A_{\rm eff}\approx13000\cm^2$ @ 1.5 keV), 
but current plans call for angular resolution
$\Delta\theta\approx6\arcsec$, which would probably preclude
accurate distance determination.
If the angular resolution were improved to $\sim1\arcsec$,
Constellation X would be capable of
extragalactic
distance determination to local group galaxies
using background AGNs with $ES_E\approx 10^{-13} \erg\cm^{-2}\s^{-1}$.
%data from http://heasarc.gsfc.nasa.gov/docs/xmm/about\_why.html

\section{\label{sec:discussion and summary}
         Discussion and Summary}

Current distance uncertainties to M31 are of order 10\%.
Bonanos et al.\ (2003) are undertaking to use detached eclipsing
binaries in M31 to establish the distance with an anticipated accuracy
of 5\%.
We have shown above that a \Chandra\ program of extended observations of
the field around the background X-ray source 5C 3.76 has the potential
to allow the distance to M31 to be determined to an absolute accuracy of
$\sim1\%$, for a 10~Ms observing campaign.
In the course of an observing campaign, we estimate a 90\% probability
that a distance estimate accurate to $\sim\pm10\%$ would emerge
after 3~Ms (see Table \ref{tab:distance uncertainties}); the distance
estimate continues to improve as more data accumulates.
We caution that our observing time 
estimates are based on assumed variability properties
for 5C 3.76; further observations to characterize its variability
should be undertaken.  In addition, monitoring of 5C 3.76 would
allow a distance-determination campaign to be ``triggered'' to begin
at a time when the source undergoes an outburst (as opposed to our
simulations, which started at random times). 

A 10~Ms observing campaign would
constitute a major
commitment of \Chandra\ observing time, but would allow determination
of the distance to M31 to unprecendented absolute accuracy.
There do not appear to be any other methods capable of determining
extragalactic distances to such precision
in the foreseeable future.
A precision determination of the distance to M31 would be of great importance
to calibrate the luminosities of stars, and to allow accurate
determination of the Hubble constant.
The proposed observing campaign would, of course, also provide a census
of the X-ray source population in this region of M31
down to lower luminosities than
current surveys, as well as information on variability of the point
sources.

To summarize our principal conclusions:
\begin{enumerate}
\item Realistic X-ray scattering properties for 
Milky Way interstellar dust
have been used
to estimate the strength of X-ray scattering halos around
distant X-ray point sources seen through foreground galaxies.
\item The X-ray variability properties of AGNs and QSOs are reviewed,
and a method for generating simulated light curves is developed
(Appendix \ref{app:light_curve}).
\item We present a simple method 
to use 
X-ray observations of a background AGN and the scattered X-ray halo
around it to directly determine the distance to a foreground
galaxy.
The methodology 
allows discontinuous observations to be employed.  Provided that the
fraction of the time lost to the gaps is not large,
discontinuous observations will be almost as effective as the idealized
continuous observations simulated here.
\item 
We use simulations to 
demonstrate the feasibility of direct determination of the
distance to M31 using the background BL Lac object 5C 3.76.
For the variability properties that we have assumed, 
a 5~Ms observing campaign with 
\Chandra\ would allow the distance to M31
to be determined to $\sim4\%$ accuracy,
and the absolute distance uncertainty
could be reduced to $<$1\% with a 10~Ms campaign
(see Figure \ref{fig:M31_distance}
and Table~\ref{tab:distance uncertainties}).
\item We discuss how observations at other wavelengths -- H~I~21cm and
CO~2.6mm aperture synthesis imaging, in particular -- 
can be used to complement the
\Chandra\ observations.
\item We consider distance determination using the X-ray halo around
GRBs.  This is a viable method for determining the distance
to the LMC or SMC,
although GRB statistics suggest that sufficiently bright
GRBs occur behind the LMC at a rate of only $\sim0.1\yr^{-1}$.
\end{enumerate}
\acknowledgements

We thank
Bohdan Paczy\'nksi for encouraging this study,
Adam Dobrzycki,
Jochen Greiner, 
Albert Kong,
Frank Primini,
Krzysztof Stanek, 
and Benjamin Williams for helpful correspondence,
and
Robert Lupton for helpful suggestions and for
making available the SM software package.
This work was supported in part by
NSF grants AST-9988126 
%btd 04.07.19 add beowulf grant
and AST-0216105.

\appendix

\section{\label{app:light_curve}Synthetic AGN Light Curve}

We describe a simple procedure to generate a stochastic light curve
with statistical properties resembling X-ray light curves for AGNs.

Let $g_j$ be a sequence of 
independent random variables with zero mean and unit variance:
$\langle g_i\rangle=0$
and $\langle g_ig_j\rangle=\delta_{ij}$.
For times $t_j=t_0+jh$, let
\beq
v_j = (1-e^{-2\lambda})^{1/2}\sum_{i=-\infty}^j g_i e^{-\lambda(j-i)}
~~~.
\eeq
If we set $\lambda = h/\tau$,
then 
\beq
\langle v_i\rangle = 0
~~~,
\eeq
\beq
\langle v_i v_j\rangle = e^{-|t_i-t_j|/\tau}
~~~.
\eeq

Consider now a (positive-definite) flux given by
\beq
\label{eq:Fdef}
F(t_j) = a e^{A v_j} ~~~~~~~(a>0)~~~.
\eeq
If the $g_j$ are assumed to be gaussian random variables, then
\begin{eqnarray}
\langle g_j^n\rangle &=& 0 ~~~{\rm for}~n~{\rm odd},\\
&=& \frac{n!}{2^{n/2}(n/2)!} ~~~{\rm for}~n~{\rm even};
\end{eqnarray}
for this case one can show that
\beq
\label{eq:<F^n>}
\langle F^n\rangle = a^n e^{n^2A^2/2} ~~~~~{\rm for~}n=0,1,2,3,...
\eeq
The r.m.s. fractional variance
\beq
F_{\rm var}\equiv
\frac{\left[\langle F^2\rangle -\langle F\rangle^2\right]^{1/2}}
{\langle F\rangle}
=
\left[e^{A^2}-1\right]^{1/2}
~~~,
\eeq
so that the coefficient $A$ can be obtained from $F_{\rm var}$:
\beq
\label{eq:A}
A = \left[\ln(1+F_{\rm var}^2)\right]^{1/2}
~~~.
\eeq
From (\ref{eq:<F^n>}) and (\ref{eq:A}) we see that the coefficient
$a$ in (\ref{eq:Fdef}) is
\beq
a = \langle F\rangle e^{-A^2/2} = 
\frac{\langle F\rangle}{(1+F_{\rm var}^2)^{1/2}}
~~~.
\eeq

The variability can be described by the structure function
\beq
{\rm SF}(\tau) \equiv \langle [F(t)-F(t-\tau)]^2\rangle
\eeq
or the power spectral density
\beq
{\rm PSD}(f) \equiv \frac{2 (\Delta t)^2}{ \Tobs} |\tilde{F}(f)|^2
~~~,
\eeq
where $\tilde{F}(f)$ is the discrete fourier transform of $F(t)$,
$\Tobs$ is the total time span observed, and $\Delta t$ is the
time between measurements.

\begin{figure}[t]
\centerline{\epsfig{
       file=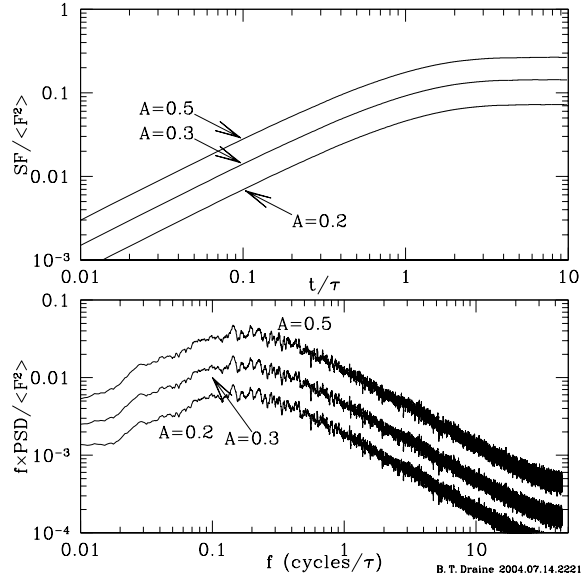,
       width=3.0in,
       angle=0}}
\caption{\label{fig:SF_and_PSD}\footnotesize
       Structure function and PSD for synthetic light curves
       for different values of $A$.
       }
\end{figure}

In Figure \ref{fig:SF_and_PSD} we show the structure function and power
spectral density obtained from these light curves.
The dimensionless quantity $f\times {\rm PSD}(f)/\langle F\rangle^2$ 
peaks at $f_p\approx 1/2\pi\tau$.
For $A=0.3$ the maximum of $f\times {\rm PSD}(f)/\langle F\rangle^2$ 
is $\sim0.015$; this is
in agreement with the maxima of $f\times {\rm PSD}(f)$ 
for the 6 Seyfert galaxies
observed by Markowitz et al.
The peak frequency $f_p\ltsim 10^{-7}\Hz$ for
Fairall 9, NGC 5547 and NGC 3516, $\sim10^{-7}\Hz$ for NGC 4151, and
$\sim 10^{-6}\Hz$ for Ark 564 and NGC 3783.

\section{Choice of Annuli
         \label{app:annuli}}

Ideally, annuli would be sufficiently narrow that the variation in
time delay across an annulus would be a fraction $\epsilon\ll 1$ 
of the time $\tau$
characterizing fluctuations of the X-ray source.
However, the minimum annulus thickness is set by the
angular resolution $\Delta\theta$ of the X-ray imager.
Thus we take the outer boundaries to be
\begin{eqnarray}
\psi_j 
&=&  \psi_1 j^{1/2}
~~~ {\rm for}~ j\leq j_c\\
&=& \psi_1 j_c^{1/2} + (j-j_c)\Delta\theta
~~~ {\rm for}~ j \geq j_c
\\
\psi_1 &\equiv& 
\left(\frac{2\epsilon c \tau}{D}\right)^{1/2}
=
2.88\arcsec 
\left(\frac{\epsilon}{.01}\right)^{1/2}
\left(\frac{\tau}{10^6\s}\right)^{1/2}
\left(\frac{\rm Mpc}{D}\right)^{1/2}
\\
j_c &=& 
{\rm nint}\left[\frac{\epsilon c \tau}{2D\left(\Delta\theta\right)^2}\right]
= {\rm nint}
\left[
2.07
\left(\frac{1\arcsec}{\Delta\theta}\right)^2
\left(\frac{\epsilon}{.01}\right)
\left(\frac{\tau}{10^6\s}\right)
\left(\frac{\Mpc}{D}\right)
\right]
\end{eqnarray}
where ${\rm nint}(x)$ is the integer nearest to $x$.

Note that annuli $j\leq j_c$ are uniformly spaced in time delay $\delta$.

\section{\label{app:simulation}
         Simulation}

Our simulated observating campaigns were created as follows.
We adopt a value for the time-averaged AGN count rate 
$\langle F\rangle \Aeff$,
the AGN correlation time $\tau$, the total exposure time $\Tobs$,
the scattering optical depth
$\tau$ (assumed uniform), and the time delay coefficient $\alpha$, where
the extra light travel time for a scattered photon from angle $\theta$
is $\alpha\theta^2$.

We assume the observing campaign to capture
$K$ images containing the AGN, 
each for an exposure time $\Tobs/K$, where $K=100$.
For simplicity, we assume that there are no time gaps between the images.
For an assumed value of the time-averaged point source count rate
$\langle F\rangle$ and correlation time $\tau$,
we compute a random realization of the AGN light curve $F(t)$
using eq.\ (\ref{eq:Fdef}), using short time steps $h\ll \tau$.

The simulated images are produced as follows:
The point source counts $N_k$ are calculated by
randomly drawing from a Poisson distribution with expectation
value $\int F(t)dt$ integrated over the exposure $k$.

Because the time delay can vary significantly across a single annulus,
we divide each annulus $j$ into a large number of subannuli.
Each subannulus has a magnification calculated using
equation (\ref{eq:magnification}), with the boundaries $\psi_j$
replaced by the boundaries of the subannulus.
We assume the background $\Ibg$ to be uniform across the image.
Each subannulus has an expected number of counts during the exposure
time $\Delta t_k$; we sum to obtain $H_{jk}^0$, the expected number of counts
for annulus $j$ during exposure $k$.
We then draw the ``observed'' counts $H_{jk}$ randomly 
from a Poisson distribution
with expectation value $H_{jk}^0$.

The simulated counts $N_k$ and $H_{jk}$ now contain Poisson ``noise''.
We take $N_k$ and $H_{jk}$ and use the method described in
\S\ref{sec:Distance Determination}
to estimate the background $\Ibg$ and time delay coefficient $\alpha$.

\section{\label{app:statistical errors}
         Statistical Errors}

Let $m_j^0$ be the true magnification of annulus $j$,
let the point source
have expected number of counts $N_k^0$ in time bin $k$, and
let the true background intensity be $\Ibg^0$.
Recall that the exposure fraction 
$f_{jk}$ is the fraction of the time $\Delta t_k$
for which annulus $j$ was observed.

Suppose we have observed counts $H_{jk}$ in annulus $j$ and time bin $k$.
For the true time delay parameter $\alpha_0$, we proceed
(as discussed in \S \ref{sec:Distance Determination}) 
to construct the
halo light curve $G_k(\alpha_0)$.
Let
\beq
A_k(\alpha) \equiv \Aeff \sum_j \Omega_j f_{jk}(\alpha)\Delta t_k
~~~.
\eeq
For each time bin $k$, the observed 
$G_k(\alpha_0)$ differs from the
expected value $[M_k^0 N_k^0 + \Ibg^0 A_k(\alpha_0)]$ 
due to Poisson fluctuations.
The expectation values $N_k^0$ are not known, but the $N_k$ are observed.
The true annular magnifications $m_j^0$ are not known; 
the estimated annular magnifications
$m_j(\alpha_0,\Ibg^0)$ and cumulative magnifications 
$M_k(\alpha_0,\Ibg^0)$
are obtained from the observed $N_k$ and the observed
$G_k(\alpha_0)$ using
eq.\ (\ref{eq:m_j}, \ref{eq:M_k}).

Assuming that we have guessed the correct value of $\Ibg=\Ibg^0$,
we wish to calculate the expectation value
$\langle
[G_k(\alpha_0) - \Ibg^0 A_k(\alpha_0) - 
M_k(\alpha_0,\Ibg^0) N_k]^2\rangle$
for $1\leq k\leq K$.
Let 
\begin{eqnarray}
M_k^0 &\equiv& \sum_{j=1}^J m_j^0 f_{jk}(\alpha_0)
\\
b_{jk}^0 &\equiv& \Ibg^0 \Aeff \Omega_j f_{jk}(\alpha_0) \Delta t_k
\end{eqnarray}
Then the observed $N_k$ and $H_{jk}$ are
\begin{eqnarray}
N_k &=& N_k^0 + \nu_k
\\
H_{jk} &=& m_j^0 f_{jk} N_k^0 + b_{jk}^{0} + x_{jk}
\end{eqnarray}
where $\nu_k$ and $x_{jk}$ are independent random variables.
The $N_k$ and $H_{jk}$ 
obey Poisson statistics, thus
\begin{eqnarray}
\langle \nu_k\rangle&=&\langle x_{jk}\rangle=0
\\
\langle \nu_k^2\rangle&=&
\langle \nu_k^3\rangle =N_k^0
\\
\langle \nu_k^4\rangle&=&3(N_k^0)^2+N_k^0
\\
\langle x_{jk}^2\rangle &=& m_j^0f_{jk}N_k^0+b_{jk}^0
\end{eqnarray}
The magnifications $m_j(\alpha_0,\Ibg^0)$ 
are estimated using eq.\ (\ref{eq:m_j}).
Expanding in powers of $[\sum_{k=1}^K f_{jk}N_k^0]^{-1}$,
\beq \label{eq:m_j estimate}
m_j = \frac{\sum_{k=1}^K H_{jk}-b_{jk}^0}{\sum_{k=1}^K f_{jk}N_k }
\approx m_j^0 + w_j^0\sum_{k=1}^K x_{jk} 
- m_j^0 w_j^0 \sum_{k=1}^K f_{jk}\nu_k
~~~,
\eeq
where
\beq
w_j^0 \equiv \left[\sum_{k=1}^K f_{jk}N_k^0\right]^{-1}
\propto \frac{1}{N}
~~~.
\eeq
We find 
\begin{eqnarray}
\langle\left[G_k - \Ibg^0 A_k - M_k N_k\right]^2\rangle
&\approx&
N_k^0 M_k^0 (1+M_k^0)
+ B_k^0
\nonumber
\\
&&
\nonumber
-2\left[(N_k^0)^2+(N_k^0)^2 M_k^0+N_k^0M_k^0\right]S_k^0
\\
\nonumber
&&+\left[4(N_k^0)^2+N_k^0\right](S_k^0)^2
\\
&&
+\left[(N_k^0)^2+N_k^0\right](T_k^0+U_k^0)
-2N_k^0V_k^0
\label{eq:<psi^2>}
~~~,
\end{eqnarray}
where 
\begin{eqnarray}
B_k^0 &\equiv& \sum_{j=1}^J b_{jk}^0
\\
S_k^0 &\equiv& \sum_{j=1}^J m_j^0 w_j^0 f_{jk}^2
% ----------------
\sim \frac{M_k}{N}
% ----------------
\\
T_k^0 &\equiv&
\sum_{j=1}^J \left(w_j^0 f_{jk}\right)^2
\sum_{\ell=1}^K \left(m_j^0 f_{j\ell} N_\ell^0  + b_{j\ell}^0\right)
% ----------------
\sim \frac{J}{N^2} \left(\frac{M_kN}{J}+\frac{B}{J}\right)
\sim
\frac{M_k}{N} + \frac{B}{N^2}
% ----------------
\\
U_k^0 &\equiv&
\sum_{\ell=1}^K N_\ell^0
\left(
\sum_{j=1}^J m_j^0 w_j^0 f_{jk}f_{j\ell}
\right)^2
% ---------------
\sim N \left( \frac{M_k}{N} \right)^2 
\sim \frac{M_k^2}{N}
% ---------------
\\
V_k^0 &\equiv&
\sum_{j=1}^J w_j^0 f_{jk} b_{jk}^0
% ---------------
\sim \frac{B}{KN}
% ---------------
\end{eqnarray}
where we have indicated the
dependences on $M_k$, $N\equiv \sum_k N_k^0$, 
and $B\equiv\sum_k B_k^0$.
Eq.\ (\ref{eq:<psi^2>}) is based on an expansion in powers of 
$(\sum_k f_{jk}N_k)^{-1}\sim J/N$, 
and therefore is applicable only if individual annuli are 
large enough that $\sum_k f_{jk}N_k^0\gg 1$.
Since $M_k^0 \ll 1$, 
in eq.\ (\ref{eq:<psi^2>}), the terms in $(S_k^0)^2$ 
and $U_k^0$ are normally negligible.
The leading order terms are 
\begin{eqnarray}
N_k^0 M_k^0 &\sim& \frac{NM}{K}
\\
B_k^0 &\sim& \frac{B}{K}
\\
(N_k^0)^2 S_k^0 &\sim& \frac{N^2 M}{K^2}
\\
(N_k^0)^2 T_k^0 &\sim& \frac{NM}{K^2} + \frac{B}{K^2}
\\
N_k^0 V_k^0 &\sim& \frac{B}{K^2}
\end{eqnarray}
It is also useful to estimate the dispersion of the estimated magnifications
$m_j$ around the true values $m_j^0$.
From eq.\,(\ref{eq:m_j estimate}):
\beq \label{eq:m_j dispersion}
\langle \left(m_j-m_j^0\right)^2\rangle \approx
m_j^0 \left(w_j^0\right)^2
\sum_k f_{jk}N_k^0
+
\left(w_j^0\right)^2 \sum_k b_{jk}^0
+
\left(m_j^0w_j^0\right)^2 \sum_k f_{jk}^2 N_k^0
~~~.
\eeq

\end{document}